\documentclass{WileyMSP-template}
\usepackage{indentfirst}
\usepackage{amsmath}
\usepackage{placeins}
\usepackage{ragged2e}

\begin{document}

\pagestyle{fancy}
\rhead{\includegraphics[width=2.5cm]{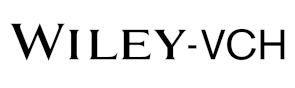}}

\title{Photonic Doping of Epsilon-Near-Zero Bragg
Microcavities}

\maketitle
\justifying

\author{Ali Panahpour*}
\author{Jussi Kelavuori}
\author{Mikko Huttunen}

\dedication{
}

\begin{affiliations}
Address:\\Photonics Laboratory, Physics Unit, Tampere University, FI-33014 Tampere, Finland

\medskip

Email Address: ali.panahpour@tuni.fi

\end{affiliations}
\medskip

\begin{abstract}

Epsilon-near-zero (ENZ) photonics provides a powerful route to extreme dispersion engineering, strong field confinement, and unconventional wave phenomena. A closely related concept is \textit{photonic doping}, where subwavelength nonmagnetic dielectric materials embedded in ENZ media enable exotic responses such as perfect-magnetic-conductor behavior and simultaneous epsilon- and mu-near-zero states. However, photonic doping has remained limited to microwave and far-infrared regimes due to the intrinsic losses of optical ENZ materials.
Here, photonic doping is demonstrated at optical frequencies by embedding a periodic array of dielectric Mie resonators into an ultralow-loss, all-dielectric ENZ platform based on near-cutoff Bragg microcavities. The resulting structures support spectrally isolated, quasi-singular coupled Bragg--Mie resonances spanning electric and magnetic multipolar orders and their overtones. These modes exhibit effective near-zero-index dispersion with fields confined either within or between the nanoparticles.
A representative \(14\,\mu\mathrm{m}\)-scale doped structure exhibits quality factors approaching \(10^{4}\) and magnetic-dipole Purcell enhancements exceeding \(5\times10^{3}\) in the near-infrared.
The demonstrated platform elevates the photonic doping from a microwave-only concept to a fully optical, low-loss, and multipole-resolved platform, enabling ultra-narrowband Mie-like resonances, enhanced magnetic-light interactions, and new opportunities in multipolar-selective spectroscopy and lasing, low-threshold nonlinear optics and efficient single-photon emission. 

\end{abstract}

\section{Introduction}
Controlling light at the nanoscale is central to nanophotonics and relies fundamentally on tailoring light–matter interactions. Maxwell’s equations dictate that optical response is set by the permittivity ($\varepsilon$) and permeability ($\mu$), which together define the refractive index $n$. 
Advances in materials science and nanofabrication now enable precise spectral and spatial control of $\varepsilon$, $\mu$, and $n$, unlocking new light–matter interaction regimes.
Among the most intriguing is the field of near-zero-index (NZI) photonics \cite{Liberal2017,Lobet2023}, encompassing materials in which the effective refractive index approaches zero due to vanishing $\varepsilon$, $\mu$, or both, commonly referred to as epsilon-near-zero (ENZ), mu-near-zero (MNZ), and epsilon-and-mu-near-zero (EMNZ) media. In NZI regimes, the optical wavelength becomes effectively infinite, the phase velocity diverges, and the wavevector tends to zero. However, some aspects of light--matter interaction, such as group velocity, wave impedance, and field confinement, vary significantly across NZI classes \cite{Lobet2020}, leading to distinct features and functionalities underpinning effects like perfect transmission, cloaking, super-coupling, diffraction suppression and enhanced light--matter interactions \cite{Liberal2017}.

A powerful approach to functionalize ENZ media is through \emph{photonic doping}, whereby subwavelength dielectric inclusions modify or encode new electromagnetic properties, analogously to electronic doping in semiconductors~\cite{Liberal2017b}. This concept has enabled extraordinary effects such as tunneling through ENZ media or general impedance matching \cite{Silveirinha2007,Zhao2019,Zhou2020,Yan2023}, geometry-independent radiation of antennas \cite{Li2022}, dispersion coding \cite{Zhou2022}, coherent perfect absorption~\cite{Luo2018CPA}, enhanced magneto-optical \cite{Liu2019_Photonic} and Kerr \cite{Nahvi2019} effects, ultra-sensitive switching and displacement sensing \cite{Wang2023}, strong concentration of magnetic fields \cite{Liberal2017c}, and the realization of perfect magnetic conductor (PMC) and EMNZ behavior using non-magnetic components \cite{Liberal2017b,Silveirinha2007}.
However, material losses and fabrication challenges at optical wavelengths have so far limited the photonic doping of ENZ media primarily to microwave and far-infrared frequencies~\cite{Salary2018,Zhao2019}.

The electromagnetic behavior introduced by the dielectric dopants is strongly influenced by their intrinsic resonant response. For highly symmetric geometries such as spheres or infinite cylinders, this response is described exactly by Mie theory, which provides closed-form analytical solutions for their multipolar scattering characteristics~\cite{Bohren2008}. Mie resonances correspond to intrinsic electric and magnetic multipolar eigenmodes determined by the particle size, refractive-index contrast, and geometry. 
High-index dielectric Mie resonators constitute foundational elements of all-dielectric photonics \cite{Rybin2024,Babicheva2024,Kruk2017} and meta-optics \cite{Kivshar2018AllDielectric,Staude2019_MetaOptics}, where their multipolar resonances facilitate wavefront control \cite{Liu2020Dielectric,Lin2025Nonlinear}, metasurface-based encryption \cite{Wang2025encryption}, nonlinear optical enhancement \cite{Wang2025,Shcherbakov2014}, and a range of multipole-mediated responses with advanced functionalities~\cite{Sugimoto2021,Babicheva2021,Liu2020MultipoleMultimode,Panahpour2011}. While individual Mie resonances are intrinsically radiative and therefore exhibit limited $Q$-factors, metasurfaces composed of dielectric Mie resonators can support collective modes with strongly suppressed radiative losses. In particular, periodic metasurfaces designed to support bound states in the continuum (BICs) can exhibit ultrahigh-$Q$ responses with strong Purcell enhancements. In such platforms,  BICs may originate from symmetry protection \cite{Li2019Symmetry}, destructive interference between multiple radiation channels (Friedrich–Wintgen BICs) \cite{Chern2023}, lattice anapole effects \cite{Allayarov2024}, or collective multipolar diffractive coupling mediated by the lattice \cite{Ha2018}.
The resulting BIC modes typically originate from collective interference that involves both lattice-mediated interactions and hybridized superpositions of multipolar contributions, with their composition dictated by the structure’s periodicity, symmetry, and the in-plane wavevector \cite{Hsu2016_BICReview}. 

Photonic doping of \textit{ENZ Bragg microcavities}~\cite{Panahpour2025,Kelavuori2024} offers a distinct route to achieving extreme spectral and spatial confinement of light.
In the ENZ regime, the radiative channels of embedded Mie resonators are intrinsically suppressed, giving rise to quasi-singular, strongly confined Mie-like modes. 
As we will show, this approach supports isolated, extremely high-\(Q\) resonances and strongly enhanced electric or magnetic local density of states (LDOS), arising from the interplay between the ENZ-induced suppression of radiation and the confinement of light provided by the cavity Bragg mirrors.
Such spectrally isolated, predominantly single-multipole modes are highly desirable for Purcell enhancement and unambiguous study of magnetic-dipole and higher-order transitions in atoms or quantum dots, where tailored resonant modes can selectively boost and spectrally shape specific emission pathways. 
Previous strategies for suppressing unwanted modes, including azimuthally polarized focused beams \cite{Kasperczyk2015} or engineered laser beams \cite{Das2015_BeamEngineering}, self-interference effects \cite{Noginova2009,KaraveliZia2011,Aigouy2014}, engineered plasmonic nanostructures \cite{Feng2011_MagneticPlasmonic,HeinGiessen2013,Hussain2015_OptLett}, and more recently ideal magnetic-dipole scattering based on anapole states in dielectric nanostructures and hybrid metal--dielectric geometries \cite{Feng2016,Feng2017,Zhang2019}, typically suffer from weak spectral isolation, modest Purcell enhancement, reduced quantum efficiency, and restricted tunability.

Interestingly, the well-defined parity and angular-momentum content of the isolated multipole modes in photonically doped systems also makes them effective \textit{symmetry filters}, enabling selective coupling to Bloch or cavity modes with matching symmetry while suppressing coupling to symmetry-incompatible channels. This property can be particularly useful for nonlinear processes such as second harmonic generation within the doped region, where restricting the generated harmonic to a single symmetry-allowed cavity mode provides an additional knob for improving symmetry-dependent modal overlap and phase matching, leading to more predictable and controllable output.
The strong $Q$-factor and magnetic PF enhancement in such platform is also of particular interest for studying magnetic light--matter interactions~\cite{Feng2016,Vaskin2019,Baranov2017,Sugimoto2021}, and 
magnetically driven enhanced nonlinear processes such as Kerr effect~\cite{Nahvi2019}, second ~\cite{Carletti2015} and third~\cite{Shcherbakov2014} harmonic generation, as well as Raman scattering~\cite{Obydennov2021,Dmitriev2016}.

Furthermore, the doped ENZ Bragg cavity provides a convenient external tuning knob through the angle of incidence. Changing the incidence angle of the exciting wave modifies the in-plane wave vector and the cavity resonance condition, thereby shifting the resonance frequency of a given dopant mode while leaving its internal field profile and dominant multipolar character unchanged. This enables unambiguous interference between any two chosen multipolar transitions, for example, bringing an electric-dipole-like and a magnetic-dipole-like dopant resonance into controlled spectral overlap, while keeping higher-order multipoles strongly suppressed. The coexistence of multiple radiative pathways with tunable amplitudes and frequencies therefore can make this platform ideally suited for exploring unambiguous quantum interference of decay channels, including modified spontaneous emission and coherent population trapping in nanophotonic environments.

In this work, we demonstrate photonic doping of ENZ media in the optical domain using all-dielectric Bragg-reflection microcavities operating near cutoff~\cite{Panahpour2025,Kelavuori2024}. We investigate spherical and cylindrical inclusions in ENZ backgrounds and focus on periodic arrays of dielectric nanocylinders (NCs) embedded within ENZ Bragg cavities. We show that this configuration yields EMNZ behavior and generates NZI coupled Bragg--Mie modes with significantly enhanced quality factors ($Q$) and Purcell factors (PF), surpassing both undoped cavities and isolated NC arrays. 
We identify two distinct families of NZI modes: (i) core-confined modes with strong $Q$ enhancement and fields localized inside or between the NCs, and (ii) mirror-confined modes with fields primarily distributed in the Bragg layers, achieving ultra-high $Q$ factors (${\sim}10^5$). 
Section~2 presents analytical results for Mie resonances in ENZ media, Section~3 provides numerical analysis of photonic doping in ENZ Bragg cavities, and Section~4 concludes with implications and future opportunities.

\section{Analytical Study of Mie Resonances in ENZ Media}

In the following theoretical and numerical analysis of the influence of ENZ media on Mie resonances, 
we assume an effectively lossless---a high figure-of-merit (FoM$=\varepsilon_r/\varepsilon_i$)---regime defined by 
\( \varepsilon_i \ll \varepsilon_r \ll 1 \), where \( \varepsilon = \varepsilon_r + i \varepsilon_i \) 
denotes the complex relative permittivity. 
In this limit, the FoM also closely approximates the ratio of the real to imaginary components of the complex refractive index, given by $\mathrm{FoM} = (n^2 - \kappa^2)/(2n\kappa) \approx n/2\kappa$, since $\varepsilon_\mathrm{i} \ll \varepsilon_\mathrm{r} \ll 1$ implies $\kappa \ll n \ll 1$.
This assumption is motivated by the fact that the $Q$-factors of Mie resonances are significantly enhanced only under such low-loss ENZ conditions \cite{Panahpour2025}. 

In this section, we provide an analytical demonstration of this effect for spherical and cylindrical dielectric objects. 
We begin by applying Mie theory to an infinitely long dielectric cylinder of radius $R$, illuminated by a normally incident plane wave with wavenumber
$k$. 
When the electric field is polarized perpendicular to the cylinder axis (TE polarization), the extinction efficiency \( Q_{\mathrm{ext}} \) is proportional to the real part of the scattering coefficients $a_n$ \cite{Bohren2008}, given by
\begin{equation}
a_n = \frac{x_i J_n(x_i) J_n'(x_e) - x_e J_n(x_e) J_n'(x_i)}{x_i J_n(x_i) {H_n^{(1)}}'(x_e) - x_e H_n^{(1)}(x_e) J_n'(x_i)},
\label{1}
\end{equation}
where \( J_n \) and \( H_n^{(1)} \) denote the Bessel and Hankel functions of the first kind, respectively, $n$ is an integer index labeling the angular (multipolar) order of the cylindrical harmonics
 and \(J_n'\) corresponds to the derivative of \(J_n\). 
The size parameters are defined as \( x_i = n_i k R \) and \( x_e = n_e k R \), corresponding to the arguments of the Bessel and Hankel functions inside and outside the cylinder.  
In these expressions, \(n_i\) and \(n_e\) denote the refractive indices of the cylinder and the surrounding medium, \(k\) is the vacuum wavenumber, and \(R\) is the radius of the cylinder.

In Fig.~1a, the real parts of the Mie coefficients \( a_0 \), \( a_1 \), and \( a_2 \) are plotted as functions of wavelength, representing the electric and magnetic resonances of an infinite dielectric cylinder with radius \( R = 150~\text{nm} \). The cylinder (\(n_i=3.5\)) is embedded in a medium with \(n_e=1\) and is illuminated by a normally incident plane wave whose electric field is perpendicular to the cylinder axis (see inset).
\begin{figure}[ht]
\includegraphics[scale=0.6]{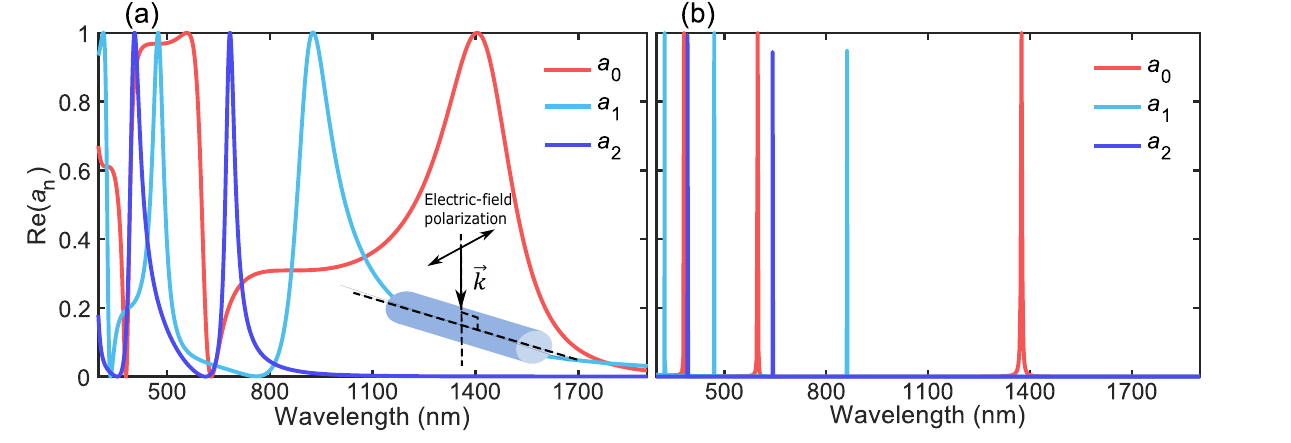}
\caption{\label{fig:1} Real parts of Mie coefficients \( a_0 \), \( a_1 \), and \( a_2 \) as functions of wavelength for an infinite cylinder of radius $R=150~$nm, $n_i=3.5$ and (a) $n_e=1$, (b) $n_e=0.1$. The inset illustrates normal incidence of a plane wave with its electric field polarized perpendicular to the cylinder axis. }
\end{figure}
Fig.~1b presents the same Mie coefficients after adjusting the refractive index of the surrounding medium to $n_e=0.1$, which results in a pronounced narrowing of the resonances. In particular, for the TE$_0$ magnetic dipole mode (Mie coefficient $a_0$), the bandwidth decreases from approximately 200 nm down to around $2\,\mathrm{nm}$.
We note that a dispersionless and purely real index $n_i=3.5$ has been used over the entire spectral range of $\lambda=300$--$1900~\mathrm{nm}$ to decouple geometry-driven modal physics from line-shaping effects due to dispersion. 
In this idealized setting, clear bandwidth narrowing takes place for all electric and magnetic  
modes in the ENZ regime. Material dispersion and loss would mainly add spectral shifts and broadening, masking but not altering the underlying mechanism.

This resonance narrowing in the ENZ regime, where \( n_e \to 0 \) and \( x_e \ll 1 \), can be analytically justified by applying Eq.~\eqref{1} and using the small-argument approximations for the functions \( J_n(x_e) \), \( J_n'(x_e) \), \( H_n^{(1)}(x_e) \), and \( {H_n^{(1)}}'(x_e) \). As demonstrated in Supporting Information (SI) document, Sec.~A, 
this procedure yields the resonance condition 
$J_n(x_i) = 0$ (consistent with the results shown in Fig.~1b), while for the resonance width ($\gamma_n$) of the TE$_n$ modes, we obtain
\begin{equation}
\gamma_n \propto \frac{1}{2(n+1)!} \left[ x_e^2 J_n'(x_i) \right] \left( \frac{x_e}{2} \right)^{n+1},
\end{equation}
which vanishes in the limit \( x_e \to 0 \).

\begin{figure}[ht]
\includegraphics[scale=0.6]{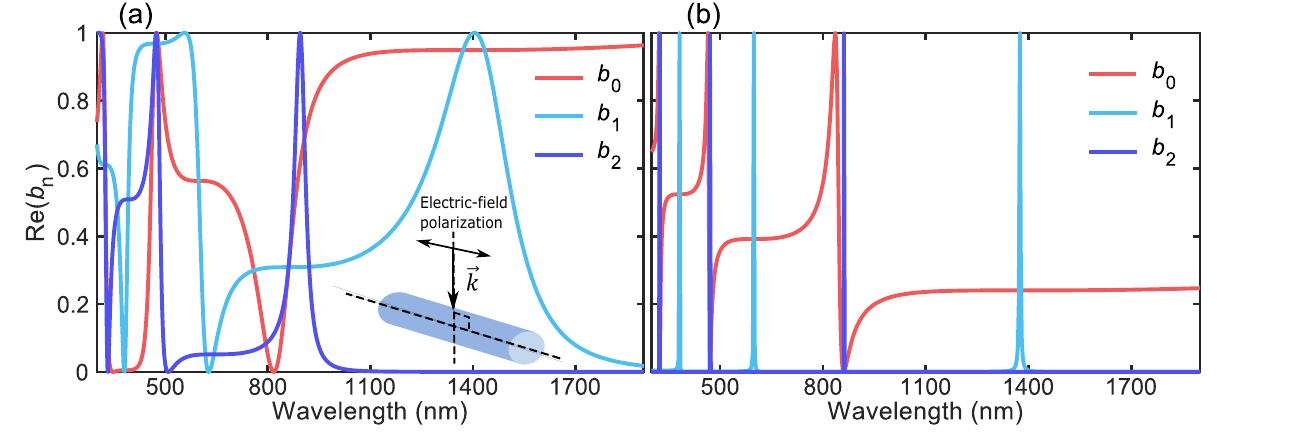}
\caption{\label{fig:2} Real parts of Mie coefficients \( b_0 \), \( b_1 \), and \( b_2 \) as functions of wavelength using the same parameters as in Fig.~1 and for (a) $n_e=1$, (b) $n_e=0.1$. }
\end{figure} 

For the case of TM modes, where the electric field is polarized along the cylinder axis (shown in the inset of Fig.~2a), the extinction efficiency is proportional to the real part of the scattering coefficients $b_n$, expressed as:
\begin{equation}
b_n = \frac{x_e J_n(x_i) J_n'(x_e) - x_i J_n(x_e) J_n'(x_i)}{x_e J_n(x_i) {H_n^{(1)}}'(x_e) - x_i H_n^{(1)}(x_e) J_n'(x_i)}.
\end{equation}

The real parts of the Mie coefficients \( b_0 \), \( b_1 \), and \( b_2 \) are plotted in Fig.~2. 
Again, significant narrowing of resonances takes place when moving from $n_e=1$ to $n_e=0.1$.
By using similar small-argument approximations for Bessel and Hankel functions, it can be shown (SI, Sec.~A) that the resonance condition for TM$_n$ modes with $n>0$ is satisfied when:
\begin{equation}
\frac{n}{2} J_n(x_i) + x_i J_n'(x_i) = 0.
\end{equation}
However, as discussed in SI, Sec.~A, the TM$_0$ mode which corresponds to a cylindrically symmetric electric monopole field distribution, does not support a true resonance, as can also be inferred from the red curve in Fig.~2b.
For the resonance width of the TM$_n$ modes with $n>0$ we get:
\begin{equation}
\gamma_n \propto \frac{1}{n!}
\left[ \frac{n}{2} J_n(x_i) - x_i J_n'(x_i) \right] \left( \frac{x_e}{2} \right)^n,
\end{equation}
which vanishes in the ENZ regime as \( x_e \to 0 \), indicating strong suppression of radiative losses and enhanced confinement.

The $Q$-factor enhancement and slight blue shift of Mie resonances in the ENZ regime, as demonstrated in Figs.~1 and 2, can be attributed to the extremely high impedance contrast between the dielectric NC and its ENZ background. This contrast suppresses scattering or radiative losses, narrowing the resonance bandwidth, while simultaneously inhibiting field penetration into the surrounding medium. The resulting strong field confinement reduces the effective mode volume, making the particle behave like a smaller optical cavity and shifting the resonance to shorter wavelengths \cite{Varghese2025}.

Similar effects can be observed for a spherical particle of radius $R$, characterized by the size parameters $x_i = n_i k R$ and $x_e = n_e k R$, corresponding to the interior and exterior of the sphere, respectively.
When the sphere is illuminated with a plane wave with wavenumber $k$, the Mie coefficients for the electric (\( a_n \)) and magnetic (\( b_n \)) multipole modes can be written in terms of the spherical Bessel functions $j_n(x)$ and 
$h_n^{(1)}(x)$ or equivalently the Riccati–Bessel functions \( \psi_n(x) = x j_n(x) \) and \( \xi_n(x) = x h_n^{(1)}(x) \) as:

\begin{equation}
a_n = \frac{x_i \psi_n(x_i) \psi_n'(x_e) - x_e \psi_n(x_e) \psi_n'(x_i)}{x_i \psi_n(x_i) \xi_n'(x_e) - x_e \xi_n(x_e) \psi_n'(x_i)},
\end{equation}
\begin{equation}
b_n = \frac{x_e \psi_n(x_i) \psi_n'(x_e) - x_i \psi_n(x_e) \psi_n'(x_i)}{x_e \psi_n(x_i) \xi_n'(x_e) - x_i \xi_n(x_e) \psi_n'(x_i)}.
\end{equation}

In SI, Sec.~B, it is shown analytically that in the ENZ regime (\( x_e \to 0 \)), the $Q$-factors of all electric and magnetic dipolar and higher-order resonances are enhanced. To illustrate this effect, the real parts of Mie coefficients,  corresponding to electric dipole ($a_1$), electric quadrupole ($a_2$), magnetic dipole ($b_1$) and magnetic quadrupole ($b_2$) resonances of a  spherical nanoparticle (NP) of radius $R=85~$nm and refractive index $n_i=3.4$ are plotted for two cases of $n_e=1$ and $n_e=0.1$, shown in Figs.~3a and 3b, respectively. 
These results extend the prediction of Ref.~\cite{Tagviashvili2010} to the electric dipolar and higher-order Mie resonances of spherical dielectric NPs.

\begin{figure}[ht]
\includegraphics[scale=0.6]{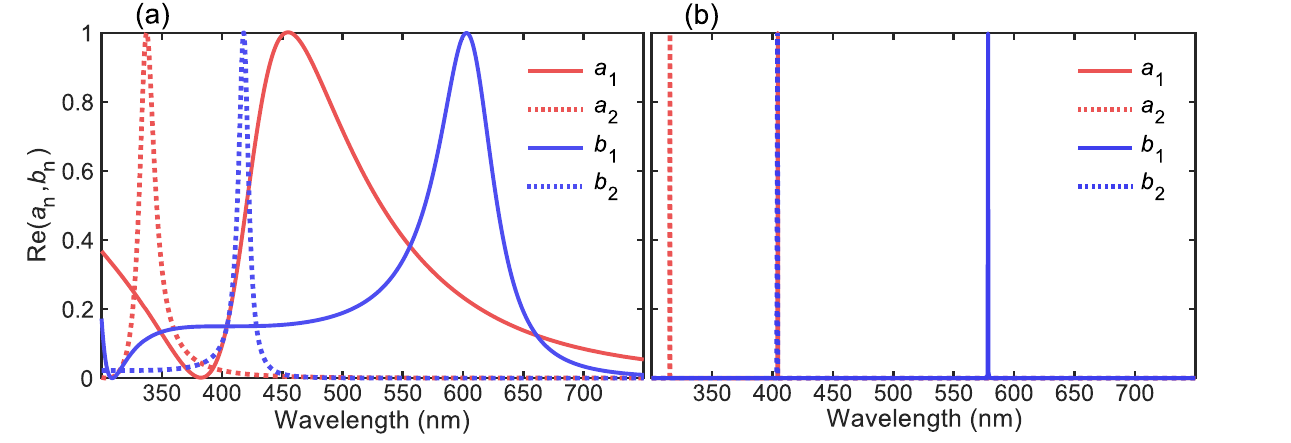}
\caption{\label{fig:3} Real parts of Mie coefficients as functions of wavelength for a spherical NP of radius 
$R=85$~nm and refractive index $n_i=3.4$, inside a medium with refractive index of (a) $n_e=1$, (b) $n_e=0.1$. }
\end{figure}

\section{Numerical Study of Photonic Doping in ENZ Bragg Cavities}

As noted earlier, most demonstrations of photonic doping in ENZ media have been carried out at microwave frequencies, typically using near-cutoff metallic or virtual PEC waveguides to emulate ENZ behavior. In such structures, even a single dielectric inclusion can strongly modify the effective electromagnetic response, giving rise to PMC- or EMNZ-like behavior \cite{Liberal2017b}.  
By contrast, analogous effects do not clearly emerge when a single NP is embedded in the core of an ENZ Bragg cavity. In this geometry, a periodic array of NPs is required to observe the photonic-doping features. This difference originates from the distinct boundary conditions imposed by PEC and Bragg mirrors. In near-cutoff PEC waveguides, the walls reflect all field components compatible with the guided-mode dispersion, so that a single dielectric inclusion perturbs essentially a single one-dimensional channel and can support a sharp photonic-doping resonance. In an ENZ Bragg cavity, however, the Bragg mirrors provide high reflectivity only within a limited spectral and angular stop band. A single NP breaks the lateral translational symmetry of the Bragg cavity and excites a broad continuum of wave vectors in different directions, most of which lie outside this stop band and are therefore poorly reflected, leading to strong leakage rather than high-\(Q\) confinement.
Introducing a periodic NP array partially restores lateral order and supports Bloch modes with well-defined in-plane wave vector that can hybridize with the ENZ cavity mode and be efficiently confined by the Bragg mirrors. 

In the dense-array limit, strong inter-particle coupling and lattice-induced collective effects modify the Bloch-mode structure \cite{Viitanen2002} and increase radiative leakage, resulting in broader resonances and reduced \(Q\) factors even in the absence of material loss. In addition, the strongly nonuniform local excitation field in this regime can admix higher-order Mie contributions, providing further channels for radiative damping.
As the lattice period is increased, inter-particle coupling weakens while the array still supports a well-defined Bloch mode compatible with Bragg confinement. In this intermediate regime, as shown in this section, the coupled Bragg-Mie resonance red-shifts by increasing the period and the quality factor increases, as radiative leakage is minimized while strong field localization inside the NPs is maintained. When the period becomes too large, however, the NPs no longer act as an effective collective photonic dopant; the modal weight inside each NP becomes small and the dominant resonances shift to modes with fields mostly concentrated between the particles or within the Bragg stacks. The coupled Bragg–Mie mode then disappears and is replaced by other cavity resonances with weak nanoparticle participation.

To numerically investigate these effects and the photonic doping features of ENZ cavities, we examine the resonance behavior of a periodic array of infinitely long dielectric cylinders embedded in ENZ Bragg cavities. Two-dimensional \textsc{Comsol} simulations are performed to calculate the transmittance of the structure shown in Fig.~4, and the eigenmodes are obtained using the mode analysis module.

\begin{figure}[ht]
\includegraphics[scale=0.9]{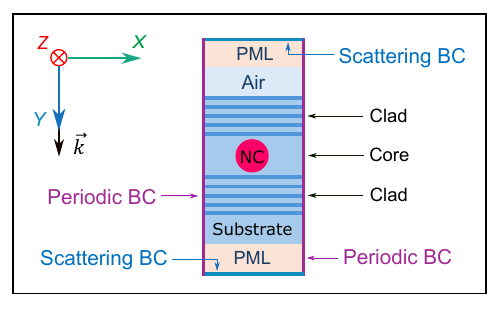}
\caption{\label{fig:4} Simulation domain of the ENZ Bragg cavity loaded with a periodic array of infinitely-long silicon cylinders of radius $R = 150~\mathrm{nm}$. The domain is enclosed by periodic and scattering boundary conditions (BCs).}
\end{figure} 

\subsection{Excitation of Transverse Electric Resonances}

Transverse Electric (TE) resonances arise when the electric field component of the incident wave is oriented perpendicular to the NC axis. Figure~5 presents the transmittance spectra of a cavity, both with and without the NC array in the core. The cavity consists of a half-wave SiO$_2$ core with refractive index of $n_{\mathrm{SiO_2}}=1.44+i10^{-6}$ \cite{Malitson1965}, sandwiched between two Bragg mirrors, each composed of 10 pairs of quarter-wave cladding layers made of SiO$_2$ and SiN ($n_{\mathrm{SiN}}=2+i10^{-7}$)\cite{Beliaev2022}. The cutoff wavelength of the cavity is set to $\lambda_c = 1500~\mathrm{nm}$. The system is doped with an array of silicon NCs with radii $R = 150~\mathrm{nm}$ and refractive index $n_{\mathrm{Si}} = 3.5+i1.38\times10^{-13}$ \cite{Polyanskiy2024}, positioned at the center of the core. The periodicity of the array is $p=500~\mathrm{nm}$ and $p=700~\mathrm{nm}$ in Figs.~5b and 5c, respectively.
\begin{figure}[ht]
\includegraphics[scale=0.75]{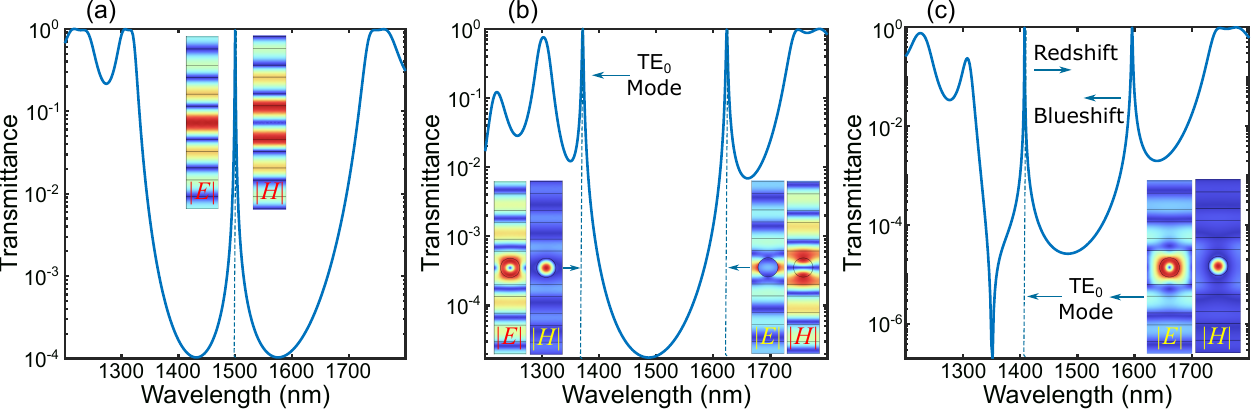}
\caption{\label{fig:5} Transmittance spectra and corresponding field profiles. (a) Bare Bragg cavity (cutoff wavelength $\lambda_c = 1500\,\mathrm{nm}$) consisting a half-wave SiO$_2$ core, sandwiched between 10 pairs of quarter-wave cladding layers of SiO$_2$ and SiN. (b,c) The same cavity loaded with arrays of infinitely-long silicon NCs, placed at the core center, with lattice periods (b) $p = 500\,\mathrm{nm}$ and (c) $p = 700\,\mathrm{nm}$. All structures are illuminated by a normally incident plane wave (along $y$-axis) with the magnetic field $H$ along $z$-axis (TE, $H_z$).
}
\end{figure} 
Figure~5a shows the transmittance spectrum and field profiles of the bare cavity mode, with a transmission peak very close to the cutoff wavelength, $\lambda_c = 1500\,\mathrm{nm}$. 
The electric field is mainly concentrated to the center of the core, while the magnetic field is localized at the core boundaries. 
Introduction of the NC array (Fig.~5b) turns the bare cavity mode into two resonant modes within the Bragg stop band, one at a longer wavelength and the other at a shorter wavelength relative to the bare-cavity cutoff. 
As demonstrated later, these correspond to NZI modes of the structure. 
The higher-energy mode, located at $\lambda = 1371~\mathrm{nm}$, is dominantly confined inside the cylinders and can be identified as the Mie TE$_0$-like mode, associated with the magnetic dipole resonance of the Mie coefficient $a_0$. 
This mode exhibits an enhanced quality factor of $Q = 626$, nearly $50$ times higher than its free-space resonance.
The lower-energy NZI mode, at $\lambda = 1623.5~\mathrm{nm}$ with a quality factor of $Q = 607$, is characterized by distinct electric and magnetic field distributions that are mainly localized outside the cylinders. Increasing the periodicity of the array preserves the general field distribution patterns of both the high- and low-energy modes. However, the $Q$-factors of the modes increase, and as shown in Fig.~5c, the lower-energy mode exhibits a blue shift while the higher-energy mode undergoes a red shift, which can extend beyond the cutoff wavelength of the bare cavity (see Fig.~6).

By increasing the periodicity to $p = 900~\mathrm{nm}$, the $Q$-factor of the TE$_0$-like mode rises to $Q = 2.32 \times 10^4$, which is about $10^3$ times higher than for corresponding free-space array and over 14 times higher than that of the bare cavity. As shown in Fig.~6, additional resonant structural modes also appear, with electromagnetic fields predominantly distributed within the Bragg layers but exhibiting exceptionally high quality factors; for example, $Q = 4.87 \times 10^5$ at $\lambda = 1459.1~\mathrm{nm}$, even higher than that of the confined TE$_0$-like mode.  
\begin{figure}[ht]
\includegraphics[scale=0.9]{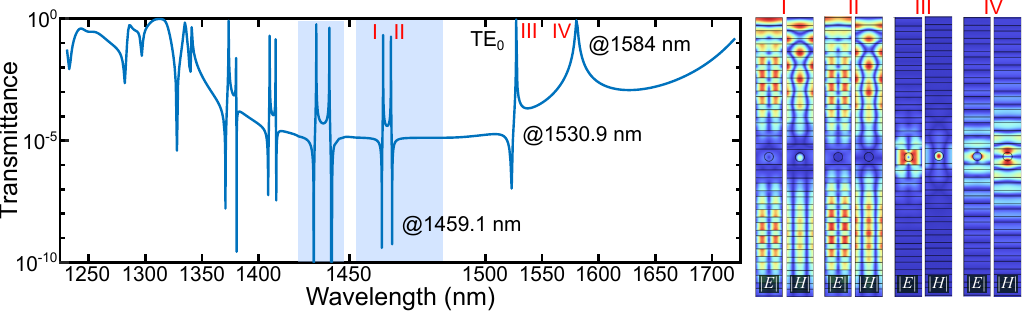}
\caption{Transmittance spectrum and field profiles of the resonant modes in the same Bragg cavity as in Fig.~5b, with a period of \( p = 900\,\mathrm{nm} \), under TE plane-wave illumination (\( \mathbf{H} \parallel z \)). To visually resolve closely spaced resonances, two wavelength intervals highlighted in blue are displayed with nonuniform horizontal scaling: the regions \(1335\text{--}1345\,\mathrm{nm}\) and \(1355\text{--}1365\,\mathrm{nm}\) are expanded by factors of 4 and 8, respectively, relative to the rest of the spectrum.}
\end{figure}

Comparison of the magnetic field distribution of the TE$_0$-like mode with that of the bare-cavity mode (Fig.~5a) reveals that enhanced field localization, together with the significant increase in $Q$-factor, produces a magnetic hot spot within the NCs, which is one of the key features of photonically doped ENZ media \cite{Liberal2017c}.

In these structures, although the period of the doped NC array modulates the transmission spectrum, leading to shifts in the resonance position and changes in bandwidth, the array periodicity by itself, when decoupled from the cavity, does not support the formation of narrowband resonances. This indicates that the emergence of high-Q resonances is primarily governed by the ENZ cavity rather than by collective lattice effects. This conclusion is supported by Fig.~7, which compares the transmittance of a TE-polarized plane wave through a single NC with that through arrays of identical NCs with varying periodicities. The curves show that arranging the NCs into an array causes the transmission peak to broaden and shift to shorter wavelengths, with the effect becoming more pronounced at smaller periods.
\begin{figure}[ht]
\includegraphics[scale=0.7]{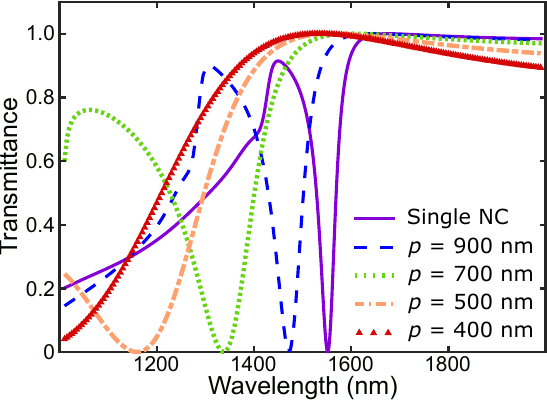}
\caption{\label{fig:7} Transmittance spectra of a single NC of radius $R=150~\mathrm{nm}$ (solid curve) and index $n_{\mathrm{Si}}$ in a SiO$_2$ matrix as well as arrays of such NCs with periods varying between 400--900~nm.
}
\end{figure}
\subsubsection{PMC and EMNZ Features in Photonically Doped Cavities}

To demonstrate the PMC and EMNZ features by photonic doping of Bragg cavities, the same structure as in Fig.~5b is considered, but with increased number of Bragg layers to $N=14$ pairs of quarter-wave layers. The transmittance spectrum of the structure is shown in Fig.~8a, together with the corresponding field profiles of the TE$_0$-like mode at $\lambda=1372.38~\mathrm{nm}$. We also computed the dispersive effective refractive index (using COMSOL’s mode analysis module), along with the effective relative permittivity and permeability around this mode, as presented in Fig.~8b. 
\begin{figure}[ht]
\includegraphics[scale=0.65]{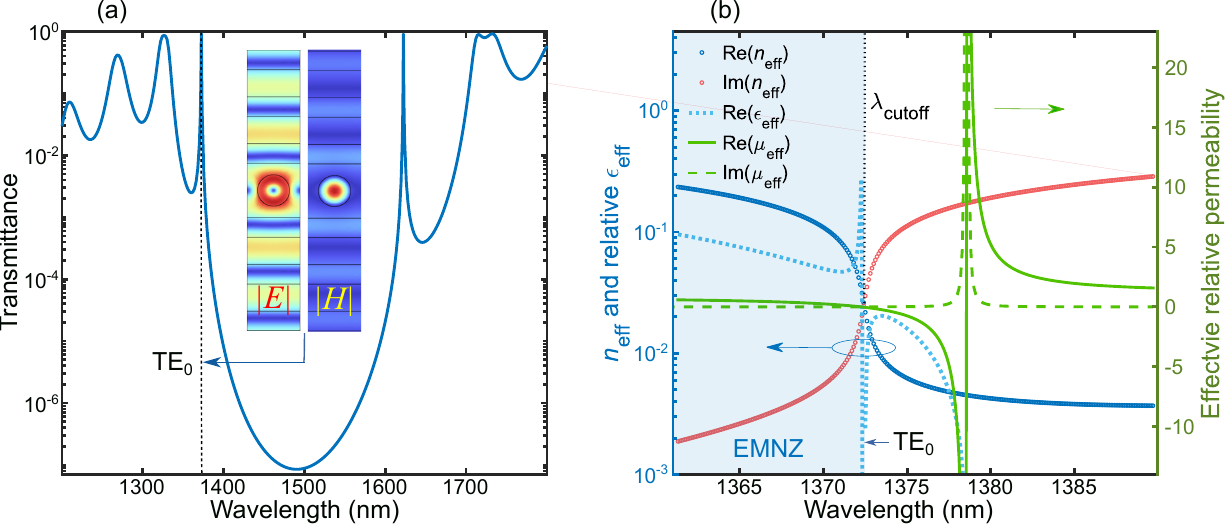}
\caption{\label{fig:8} (a) Transmittance and resonant-mode field distribution corresponding to a Bragg cavity with $N=14$ pairs of cladding layers on each side of the half-wave core and doped with an array of cylindrical NCs of radius $R=150~\mathrm{nm}$ having a period of $p=500~\mathrm{nm}$. (b) Calculated effective refractive index as well as effective relative permittivity and permeability around the wavelength $\lambda=1372.38~\mathrm{nm}$ corresponding to the TE$_0$-like resonant mode in the same NC-doped Bragg cavity. 
}
\end{figure}
In order to consistently define the effective magnetic permeability of a periodic array of inclusions, we follow the homogenization procedure introduced by Silveirinha and Engheta~\cite{Silveirinha2006}. 
For a periodic Bloch mode, the microscopic electric and polarization current density fields have the form \cite{Chremmos2015}
\begin{equation}
\mathbf{E}(\mathbf r)=\tilde{\mathbf E}(\mathbf r)\,e^{i\mathbf k_B\cdot\mathbf r}, 
\qquad 
\mathbf{J}(\mathbf r)=\tilde{\mathbf J}(\mathbf r)\,e^{i\mathbf k_B\cdot\mathbf r},
\end{equation}
where $\tilde{\mathbf E}$ and $\tilde{\mathbf J}$ are periodic functions over the unit cell and $\mathbf k_B$ is the Bloch wavevector. 
Accordingly, the microscopic magnetic dipole moment per unit length is written as \cite{Chremmos2015}
\begin{equation}
m_z \;=\; \frac{1}{2} \iint_{\text{inc}}
\Big[(x-x_c) J_y - (y-y_c) J_x\Big] \, e^{-i\mathbf k_B\cdot\mathbf r}\, \mathrm{d}A,
\end{equation}
where $J_x=-i\omega(\varepsilon-\varepsilon_{\text{host}})E_x$ and $J_y=-i\omega(\varepsilon-\varepsilon_{\text{host}})E_y$ are the polarization current densities inside the inclusion, and $(x_c,y_c)$ is the centroid of the inclusion cross-section, included to avoid origin dependence. 
The corresponding magnetization is normalized over the whole unit cell area,
\begin{equation}
M_z = \frac{m_z}{A_{\text{cell}}}, 
\qquad 
A_{\text{cell}} = \iint_{\text{cell}} \mathrm{d}A.
\end{equation}

The dephasing factor $e^{-i\mathbf k_B\cdot\mathbf r}$ is also used consistently in the evaluation of all other cell-averaged quantities, such as $\langle H_z \rangle$, ensuring that the retrieved effective permeability $\mu_\text{eff}$ is invariant under translations of the unit cell and fully consistent with Floquet--Bloch homogenization theory. However, since the mode under consideration is an NZI mode, with effective index close to zero, the Bloch wavevector is negligible and the dephasing factor can be safely ignored.

Since the direct cell average of the magnetic field $\langle H_z\rangle$ contains contributions from both the applied field and the self-field of the induced dipole, a corrected bulk magnetic field is introduced as \cite{Silveirinha2007}
\begin{equation}
H_{\text{bulk}} = \langle H_z\rangle - \frac{m_z}{A_{\text{cell}}}.
\end{equation}
The constitutive relation can then be written as
\begin{equation}
B = \mu_0 \langle H_z\rangle = \mu_{\text{eff}} H_{\text{bulk}},
\end{equation}
which leads to the following expression for the effective relative permeability:
\begin{equation}
\mu_r = \frac{\mu_{\text{eff}}}{\mu_0} = \frac{\langle H_z\rangle}{H_{\text{bulk}}}.
\end{equation}

In Fig.~8b, the real and imaginary parts of the effective index of refraction are shown by the dark-blue and red-circle curves, respectively. The real and imaginary parts of the effective relative permeability are plotted as green curves, while the effective relative permittivity, calculated from the relation $\varepsilon_{\mathrm{eff}} = n_{\mathrm{eff}}^{2}/\mu_{\mathrm{eff}}$, is shown by the dotted light-blue curve. 

The TE$_0$-like mode resonates at $\lambda = 1372.38~\mathrm{nm}$, in the close vicinity of the cutoff. The PMC-like response appears at the wavelength where the effective permeability (solid green) exhibits a pole ($\mu_r \to \infty$), corresponding to a high-impedance condition.
The EMNZ region, corresponding to wavelengths shorter than the cutoff wavelength, is highlighted in blue. In this region, both the effective permittivity and permeability are positive and smaller than unity. For wavelengths longer than the cutoff wavelength, the signs of the effective permittivity and permeability are opposite, and consequently the wave cannot propagate in this spectral range, as indicated by the rapid growth of the imaginary part of the effective index (red-circle curve).

It is interesting to note that, although the microscopic magnetic dipole moment $m_z$ and consequently the TE$_0$-like mode exhibits a resonance at $\lambda = 1372.38~\mathrm{nm}$, the effective relative permeability diverges at a slightly longer wavelength. This red shift follows directly from the retrieval formula
\begin{equation}
\mu_r = \frac{\langle H_z \rangle}{\langle H_z \rangle - M_z}, 
\qquad M_z = \frac{m_z}{A_{\text{cell}}}.
\end{equation}
While $M_z$ peaks at the intrinsic magnetic-dipole-like resonance of the inclusions, the pole of $\mu_r$ occurs when the bulk field in the denominator, $H_{\text{bulk}} = \langle H_z \rangle - M_z$,
vanishes. The relative phase and dispersive behavior of $\langle H_z \rangle$ therefore shift the divergence of $\mu_r$ to longer wavelengths than the maximum of $M_z$.

\subsection{3.2 Excitation of Transverse Magnetic (TM) Resonances}

The TM resonances of the doped Bragg cavity are excited by illumination with a normally incident plane wave with its electric field polarized along the $z$-axis. The transmittance and field profiles of TM$_1$-like modes are shown in Fig.~9. For a period of $p = 500~\mathrm{nm}$ only the TM$_1$-like mode exists, and is confined inside the NCs (Fig.~9a). 
In contrast, when the period is increased to $p = 750~\mathrm{nm}$, an additional mode emerges at shorter wavelengths (Fig.~9b). The fields of this mode are primarily confined to the regions between the NCs. When the period is further increased to $p = 900~\mathrm{nm}$, both modes undergo a redshift, approach each other, and exhibit enhanced $Q$-factors (Fig.~9c). At the same time, other modes also appear at shorter wavelengths, with their fields predominantly distributed in the Bragg mirror layers rather than within the cavity core.
\begin{figure}[ht]
\includegraphics[scale=0.8]{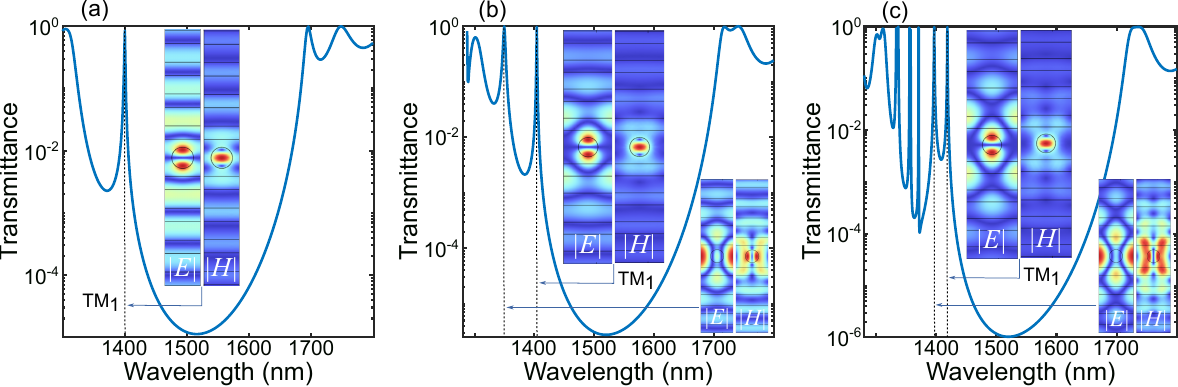}
\caption{\label{fig:9} Transmittance and the resonant-mode field distribution of a Bragg cavity with the same parameters as in Fig.~5, loaded with an array of NCs ($R=150~\mathrm{nm}$), for periods of (a) $p=500~\mathrm{nm}$, (b) $p=750~\mathrm{nm}$, and (c) $p=900~\mathrm{nm}$. All structures are illuminated by a normally incident plane wave (along $y$-axis) with the $E$-field along $z$-axis (TM, $E_z$).
}
\end{figure} 
 \subsection{3.3 Tunability of the Bragg-Mie Modes Beyond the Size Parameter}
In addition to the conventional size parameters $x_i$ and $x_e$, which depend on wavelength, particle size and the refractive indices of the particle and surrounding medium,\cite{Bohren2008} the coupled Bragg--Mie modes can be further tuned by external and structural parameters. Specifically, varying the angle of incidence (AoI) and the bare-cavity cutoff wavelength introduces additional means for controlling the resonance wavelength of the Bragg--Mie modes. These parameters enable tunability of the resonances beyond what is achievable with the size parameter alone, offering a richer degree of freedom for engineering their spectral position and field confinement. For instance, this approach can enable straightforward fine tuning and alignment of the mode resonances with the magnetic dipole resonances of doped atoms or quantum dots, facilitating enhancement or selective excitation of their intrinsic magnetic dipole transitions.

Figure~10 illustrates this tunability for both the TE$_0$-like (Fig.~10a) and TM$_1$-like (Fig.~10b) modes by showing the transmittance of a doped Bragg cavity with the same parameters as in Fig.~5, for $p = 500~\mathrm{nm}$ and different bare-cavity cutoff wavelengths of $\lambda_c = 1450~\mathrm{nm}$, $1500~\mathrm{nm}$, and $1550~\mathrm{nm}$. For both modes, the resonance redshifts for larger cutoff wavelengths.
\begin{figure}[ht]
\includegraphics[scale=0.6]{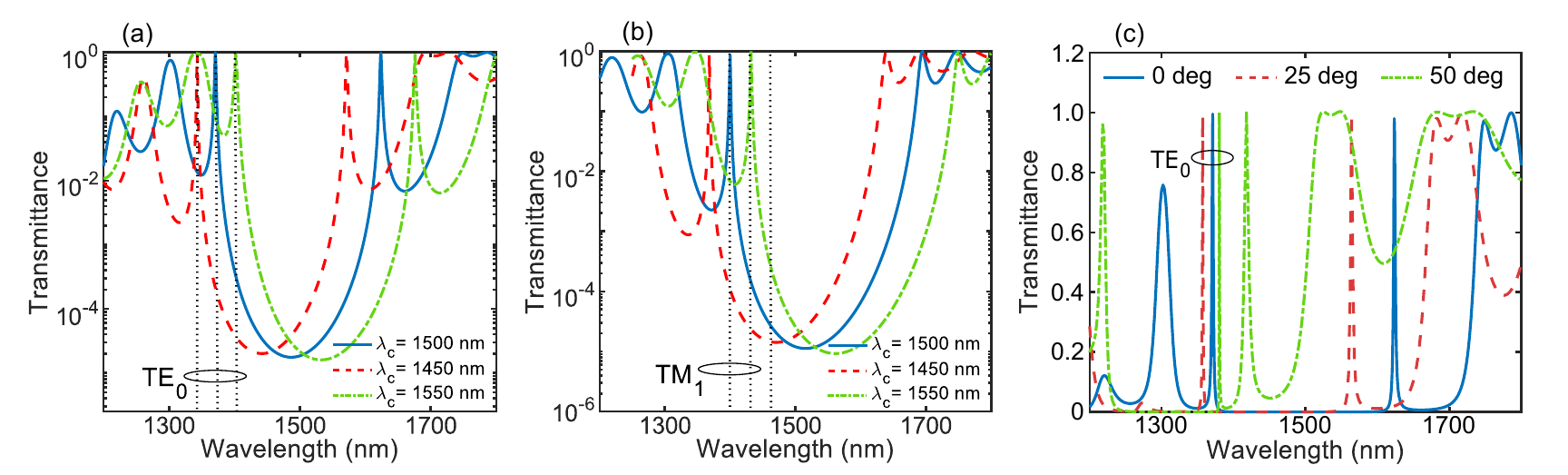}
\caption{Transmittance spectra of a doped Bragg cavity with period $p = 500\,\mathrm{nm}$ for (a) TE$_0$-like and (b) TM$_1$-like modes at different cutoff wavelengths $\lambda_c = 1450\,\mathrm{nm}$ (red dashed), $1500\,\mathrm{nm}$ (blue solid), and $1550\,\mathrm{nm}$ (green dotted). (c) TE-polarized transmittance of the same structure at $\lambda_c = 1500\,\mathrm{nm}$ for incidence angles of $0^\circ$ (solid), $25^\circ$ (dashed), and $50^\circ$ (dash-dotted).}

\end{figure} 

Figure~10c shows the TE-polarized transmittance of the same structure at $\lambda_c = 1500\,\mathrm{nm}$ for increasing AoI. The lower- and higher-energy modes shift and move closer together, illustrating an angle-tunable response; the lower-energy branch blue-shifts across the range, while the higher-energy branch shows a weaker, non-monotonic shift at larger angles.

Figure~11 shows the TM-polarized transmittance of the same structure as in Fig.~10, but with $p = 750~\mathrm{nm}$ and AoIs of $0^\circ$, $5^\circ$, and $15^\circ$. 
As seen in Fig.~10a, for non-zero AoIs, new resonances emerge with markedly high $Q$-factors, particularly at smaller angles (e.g., $Q \approx 1.5 \times 10^4$ for AoI $=5^\circ$, compared to $Q \approx 700$ for the TM$_1$-like mode). This resonance is structurally similar to the TM$_1$-like mode but exhibits a $90^\circ$ rotated field distribution. Upon increasing AoI, the mode redshifts while the TM$_1$-like mode gradually vanishes. As illustrated in Fig.~11b for an AoI of $15^\circ$, additional resonances also appear at shorter wavelengths, with their fields predominantly distributed in the Bragg mirror layers rather than in the cavity core.

\begin{figure}[ht]
\includegraphics[scale=0.78]{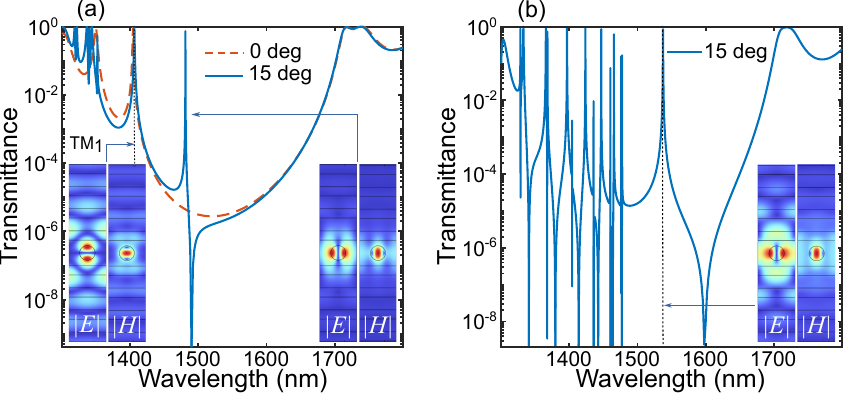}
\caption{\label{fig:11} TM-polarized transmittance of a Bragg cavity with $N=10$ pairs of cladding layers on each side of a half-wave core, designed for a cutoff wavelength of $1500~\mathrm{nm}$, and loaded with an array of cylindrical NCs ($R=150~\mathrm{nm}$) with period $p=750~\mathrm{nm}$. (a) Results for incidence angles of $0^{\circ}$ (dashed curve) and $5^{\circ}$ (solid curve). (b) Results for an incidence angle of $15^{\circ}$.}
\end{figure}
The modes discussed above---including the TE- and TM-like resonances confined inside the NCs, the modes localized between the NCs, and the high-$Q$ resonance emerging at nonzero AoI in Fig.~11a---can all be identified as high-FoM NZI modes of the doped Bragg cavity. This is confirmed by the dispersion curves shown in Fig.~12a in the wavelength range of 1300--1650~nm, together with their corresponding transmission peaks presented in Fig.~12b.

\begin{figure}[ht]
\includegraphics[scale=0.65]{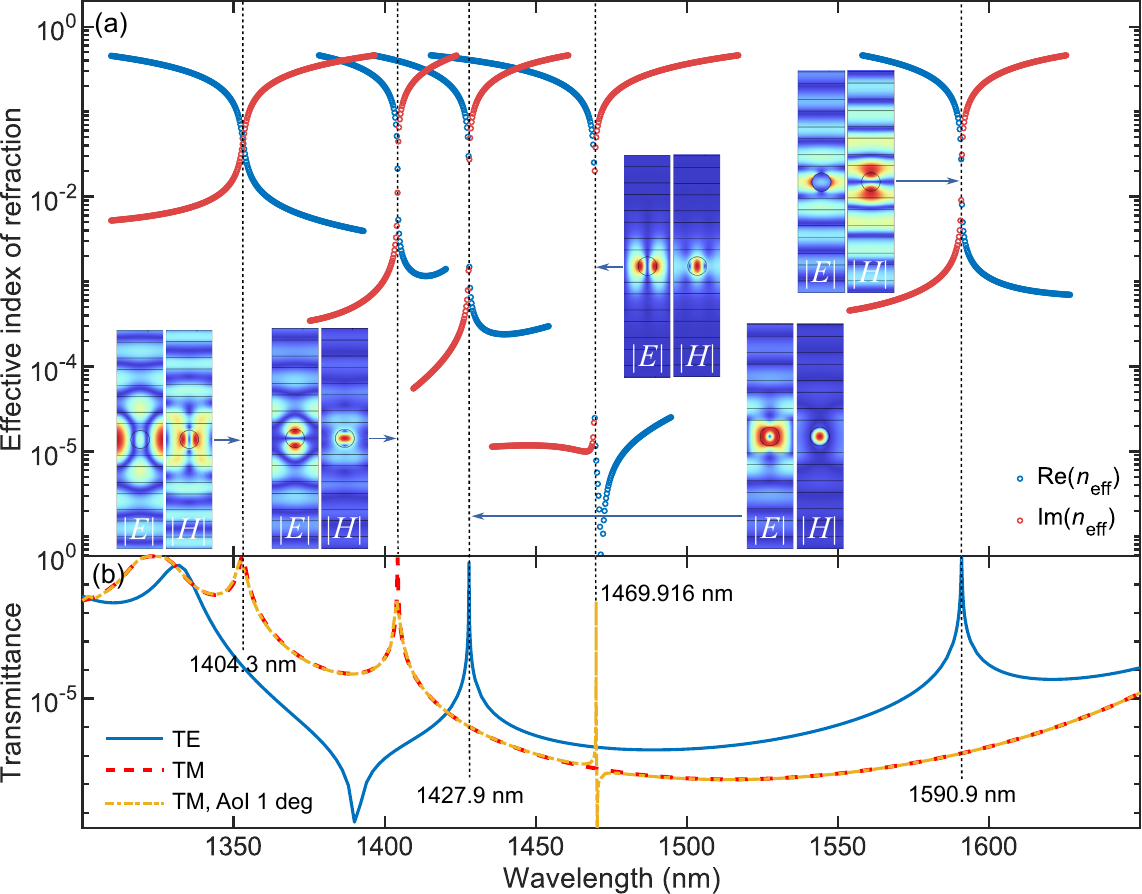}
\caption{\label{fig:12} (a) Real and imaginary parts of effective refractive indices of TE and TM modes, together with their corresponding field distributions, in a Bragg cavity with $N=14$ pairs of cladding layers on each side of a half-wave core, loaded with an array of cylindrical NCs, with a period of $p=750~\mathrm{nm}$. (b) Transmission spectrum of the same cavity for TE (solid curve) and TM (dashed/dotted curves) modes.}
\end{figure} 

\subsection*{3.4 Magnetic Purcell Factor Enhancement in NC-Doped ENZ Cavity}

To quantify the spontaneous-emission enhancement in the NC-doped ENZ Bragg cavity, we evaluate the PF and compare it with that of a single NC or an array of NCs embedded in an infinite homogeneous medium. The formulation follows the Green-tensor approach previously introduced for two-dimensional (2D) photonic systems~\cite{Qiao2011}, extended here to account for both electric and magnetic dipole emitters by means of the corresponding electric and magnetic Green functions.  

For an electric dipole $\mathbf p=p_0\,\mathbf u_p$, with orientation $\mathbf u_p$ at position $\mathbf r_0$, the normalized decay rate is given by the LDOS expression \cite{Novotny2012}
\begin{equation}
\frac{\Gamma(\omega)}{\Gamma_0(\omega)}
= \frac{6\pi}{k}\,\Im\!\left\{\mathbf u_p\!\cdot\!\mathbf G(\mathbf r_0,\mathbf r_0;\omega)\!\cdot\!\mathbf u_p\right\},
\qquad k=\omega/c .
\end{equation}
Using the relation between the Green tensor and the local electric field (SI, Sec.~C), one obtains the relation \cite{Lobet2020}
\begin{equation}
{
\frac{\Gamma(\omega)}{\Gamma_0(\omega)}
= -\,\frac{6\pi c}
{\mu_0\,\omega^{3}\,|p_0|}\,
\Re\!\left\{
\mathbf u_p\!\cdot\!\mathbf E(\mathbf r_0,\omega)
\right\}.
}
\label{eq:C-final-elec}
\end{equation}

A parallel derivation applies to a magnetic dipole $\mathbf m = m_0\,\mathbf u_m$, with orientation $\mathbf u_m$ at $\mathbf r_0$. Starting from the magnetic LDOS and invoking the relation between $\mathbf H$ and the magnetic Green tensor (SI, Sec.~C), we obtain

\begin{equation}
{\
\frac{\Gamma_m(\omega)}{\Gamma_{m,0}(\omega)}
= -\,\frac{6\pi c}{\varepsilon_0 \,\omega^{3}\,|m_0|}\,
\Re\!\left\{\mathbf u_m\!\cdot\!\mathbf H(\mathbf r_0,\omega)\right\}.\
}
\end{equation}
For our 2D cavity, the PF is evaluated by exciting the structure with either an electric or a virtual magnetic line current placed at the center of an NC in the core, oriented along $z$. The PF is then obtained from the above expressions by computing the real part of the corresponding $z$-component of the field at the source position and normalizing to its free-space value.  

Figure~13a compares the calculated PF spectra at the center of a single silicon NC in free space (dashed line), an array of 25 NCs in free space (dashed-dotted line), and the same NC array embedded in the Bragg cavity (solid line). The embedded array exhibits a nearly 35-fold Purcell enhancement relative to the isolated NC or free-space array. The corresponding electric and magnetic field distributions at resonance are shown in Figs.~13b and 13c. In these simulations, the Bragg cavity consists of $N=14$ cladding-layer pairs on each side of a half-wave core, loaded with an array of 25 silicon NCs of radius $R=150~\mathrm{nm}$ with period $p=750~\mathrm{nm}$. The cavity width is $W=18.75~\mu\mathrm{m}$.
\begin{figure}[ht]
\includegraphics[scale=0.7]{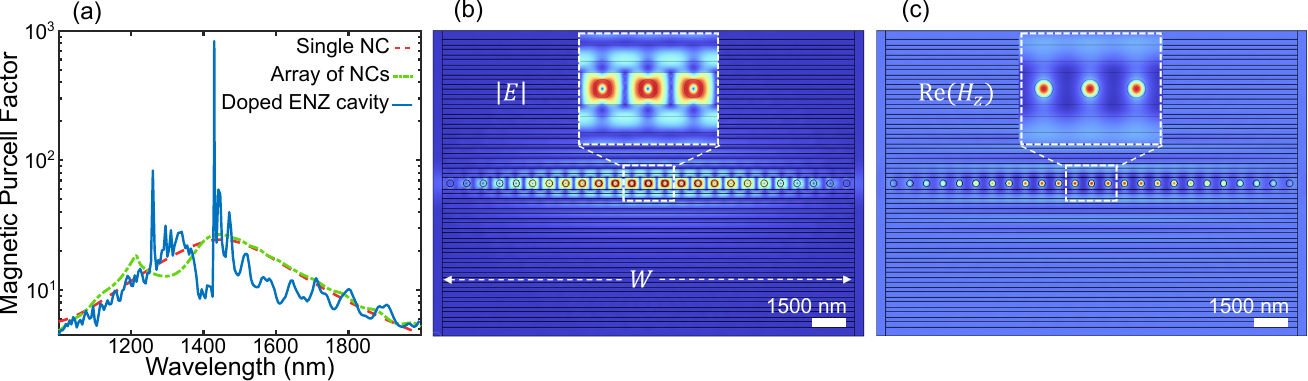}
\caption{\label{fig:13} (a) Magnetic PF for a single NC in free space, an array of 25 NCs in free space, and the same NC array embedded in the Bragg cavity. (b) Electric-field magnitude $|E|$ and (c) magnetic-field component $\mathrm{Re}(H_z)$ at the resonant wavelength. The Bragg cavity comprises of $N=14$ cladding-layer pairs on each side of a half-wave core and is loaded with an array of 25 NCs ($R=150~\mathrm{nm}$, $n=n_{\mathrm{Si}}$, $p=750~\mathrm{nm}$).}

\end{figure}
The Purcell enhancement effect can also be observed in higher-order NZI modes of the cavity, which arise from reflections at the cavity sidewalls due to imperfect impedance matching with the surrounding medium, thereby forming standing-wave patterns inside the cavity. Interestingly, by tuning the cavity width, the PF at the center of cavity, associated with these higher-order modes can be optimized. For example, as shown in Fig.~14a, for an array of 25 silicon NCs ($R=150~\mathrm{nm}$ and $p=500~\mathrm{nm}$) embedded in a Bragg cavity with $N=14$ and a width of $W=14~\mu\mathrm{m}$, the PF reaches values as high as 5600 for a higher-order NZI TE$_0$-like mode, with the corresponding magnetic field profile shown in Fig.~14c.

\begin{figure}[ht]
\includegraphics[scale=0.75]{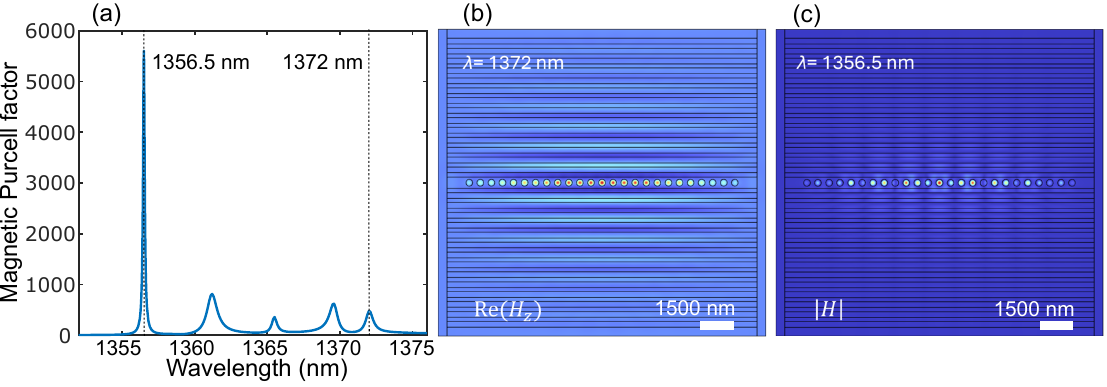}
\caption{\label{fig:14} 
Purcell enhancement in higher-order NZI modes of a Bragg cavity with $N=14$ cladding-layer pairs and width $W=14~\mu\mathrm{m}$, loaded with an array of 25 NCs ($R=150~\mathrm{nm}$) and period $p=500~\mathrm{nm}$. 
(a) Magnetic Purcell factor spectrum showing multiple resonances, with a maximum PF $\approx 5600$ at $\lambda=1356.5~\mathrm{nm}$. 
(b) Magnetic field distribution $\Re(H_z)$ at $\lambda=1372~\mathrm{nm}$ corresponding to a lowest-order cavity mode. 
(c) Magnetic field distribution $|H|$ at $\lambda=1356.5~\mathrm{nm}$, corresponding to a highest-order NZI TE$_0$-like mode responsible for the giant PF enhancement.}

\end{figure}

The electric Purcell enhancement in NC-doped Bragg cavities can be investigated for different orders of NZI TM-polarized modes by exciting the structure with electric line currents placed at the electric hot spots of the modes (see SI, Sec.~D). It should be noted that the use of electric or virtual magnetic line currents is purely a computational tool to excite the corresponding TM or TE modes; in practice, the TE- and TM-polarized modes can be excited by plane-wave illumination with the appropriate polarization.

\section{Conclusion}
It is shown that photonic doping of ENZ Bragg cavities with periodic arrays of dielectric NCs provides a robust route to engineer high-$Q$ near-zero-index modes with well-controlled electric and magnetic character. On the analytical side, we demonstrated that embedding Mie resonators in an ENZ background generically suppresses radiative losses and narrows both electric and magnetic multipolar resonances for cylindrical and spherical inclusions. Numerically, we established that near-cutoff all-dielectric Bragg cavities doped with resonant silicon NC arrays support coupled Bragg--Mie modes with effective EMNZ response, quality factors approaching $10^{4}$, and magnetic-dipole Purcell enhancements exceeding $5\times10^{3}$ for a \(14\,\mu\mathrm{m}\)-scale doped structure, far beyond what is achievable with isolated particles or undoped cavities.

The high-Q NZI modes can be selectively excited and tuned through structural and external parameters such as lattice period, cavity width, cutoff wavelength, and angle of incidence. This tunability enables access to spectrally isolated, predominantly single-multipole resonances and to controlled interference between electric- and magnetic-dipole-like channels. The high-$Q$ magnetic resonances of the NCs realize optical-frequency analogues of PMC and EMNZ behavior in a fully dielectric platform and provide large LDOS enhancements for both electric and magnetic emitters, with well-defined multipolar and polarization selectivity.

Beyond the specific configurations considered here, the concept of ENZ photonic doping is readily extendable to alternative dopant geometries, higher-order cavity modes, and three-dimensional architectures, as well as to active or nonlinear materials. We anticipate that NC-doped ENZ Bragg cavities will serve as a versatile building block for low-threshold and multipolar-selective lasing, magnetic-dipole and higher-order spectroscopy, enhanced magnetically driven nonlinear processes, and engineered quantum light--matter interactions.


\medskip
\textbf{Supporting Information} \par 
Analytical derivations of Mie resonances for cylindrical and spherical inclusions embedded in ENZ media, including the ENZ-limit behavior of TE and TM scattering coefficients and resonance linewidths. Formal definitions of electric and magnetic dipoles, dyadic Green tensors, local density of states, and Purcell factors are provided. Additional numerical results demonstrate electric Purcell enhancement in photonically-doped ENZ Bragg microcavities under TM-polarized excitation, together with corresponding near-field distributions.
Supporting Information is available from the Wiley Online Library or from the author.

\medskip
\textbf{Acknowledgements} \par 
All authors acknowledge the support of the Flagship of Photonics Research and Innovation (PREIN) funded by the Research Council of Finland (grant no.~320165). JK also acknowledges the Magnus Ehrnrooth foundation for their PhD grant.

\FloatBarrier

\bibliographystyle{MSP}
\bibliography{references}

@article{Panahpour2011,
  author  = {Panahpour, A. and Latifi, H.},
  title   = {Electromagnetic transparency and slow light in an isotropic 3D optical metamaterial, due to Fano-like coupling of Mie resonances in excitonic nano-sphere inclusions},
  journal = {Optics Communications},
  volume  = {284},
  number  = {6},
  pages   = {1701--1710},
  year    = {2011}
}

@article{Varghese2025,
author = {Varghese, Riya and Annurakshita, Shambhavee and Tamashevich, Yaraslau and Chellu, Abhiroop and Bej, Subhajit and Rekola, Heikki and Lyytikäinen, Jari and Wahl, Hanna and Schildt, Matias and Panahpour, Ali and Niemi, Tapio and Ornigotti, Marco and Karvinen, Petri and Guina, Mircea and Hakkarainen, Teemu and Huttunen, Mikko J.},
title = {Nonlinear Optical Microscopy of Semiconductor–Metal Nanocavities},
journal = {Advanced Photonics Research},
volume = {6},
number = {11},
pages = {2500114},
doi = {https://doi.org/10.1002/adpr.202500114},
url = {https://advanced.onlinelibrary.wiley.com/doi/abs/10.1002/adpr.202500114},
eprint = {https://advanced.onlinelibrary.wiley.com/doi/pdf/10.1002/adpr.202500114},
year = {2025}
}

@article{Lobet2023,
  author  = {Lobet, Michaël and Kinsey, Nathaniel and Liberal, Iñigo and Caglayan, Humeyra and Huidobro, Paloma A. and Galiffi, Emanuele and Mejía-Salazar, Jorge Ricardo and Palermo, Giovanna and Jacob, Zubin and Maccaferri, Nicolò},
  title   = {New horizons in near-zero refractive index photonics and hyperbolic metamaterials},
  journal = {ACS Photonics},
  volume  = {10},
  number  = {11},
  pages   = {3805--3820},
  year    = {2023},
  doi     = {10.1021/acsphotonics.3c00649},
  url     = {https://doi.org/10.1021/acsphotonics.3c00649}
}

@ARTICLE{Liberal2017,
  author  = {Liberal, I{\~n}igo and Engheta, Nader},
  title   = {Near-zero refractive index photonics},
  journal = {Nature Photonics},
  volume  = {11},
  number  = {3},
  pages   = {149--158},
  year    = {2017},
  doi     = {10.1038/nphoton.2017.13},
  url     = {https://doi.org/10.1038/nphoton.2017.13}
}

@article{Liberal2017b,
  author  = {Liberal, Iñigo and Mahmoud, Ahmed M. and Li, Yue and Edwards, Brian and Engheta, Nader},
  title   = {Photonic doping of epsilon-near-zero media},
  journal = {Science},
  volume  = {355},
  number  = {6329},
  pages   = {1058--1062},
  year    = {2017},
  doi     = {10.1126/science.aal4273},
  url     = {https://doi.org/10.1126/science.aal4273}
}

@article{Zhao2019,
  author  = {Zhao, Lin and Feng, Yijun and Zhu, Bo and Zhao, Junming},
  title   = {Electromagnetic properties of magnetic epsilon-near-zero medium with dielectric dopants},
  journal = {Optics Express},
  volume  = {27},
  number  = {14},
  pages   = {20073--20083},
  year    = {2019},
}

@article{Zhou2020,
  author  = {Zhou, Ziheng and Li, Yue and Nahvi, Ehsan and Li, Hao and He, Yijing and Liberal, Iñigo and Engheta, Nader},
  title   = {General impedance matching via doped epsilon-near-zero media},
  journal = {Physical Review Applied},
  volume  = {13},
  number  = {3},
  pages   = {034005},
  year    = {2020},
  doi     = {10.1103/PhysRevApplied.13.034005},
  url     = {https://doi.org/10.1103/PhysRevApplied.13.034005}
}

@article{Li2022,
  author  = {Li, Hao and Zhou, Ziheng and He, Yijing and Sun, Wangyu and Li, Yue and Liberal, Iñigo and Engheta, Nader},
  title   = {Geometry-independent antenna based on Epsilon-near-zero medium},
  journal = {Nature Communications},
  volume  = {13},
  number  = {1},
  pages   = {3568},
  year    = {2022},
  doi     = {10.1038/s41467-022-31234-0},
  url     = {https://doi.org/10.1038/s41467-022-31234-0}
}

@article{Zhou2022,
  author  = {Zhou, Ziheng and Li, Hao and Sun, Wangyu and He, Yijing and Liberal, Iñigo and Engheta, Nader and Feng, Zhenghe and Li, Yue},
  title   = {Dispersion coding of ENZ media via multiple photonic dopants},
  journal = {Light: Science \& Applications},
  volume  = {11},
  number  = {1},
  pages   = {207},
  year    = {2022},
  doi     = {10.1038/s41377-022-00885-2},
  url     = {https://doi.org/10.1038/s41377-022-00885-2}
}

@article{Silveirinha2007,
  author  = {Silveirinha, Mário and Engheta, Nader},
  title   = {Design of matched zero-index metamaterials using nonmagnetic inclusions in epsilon-near-zero media},
  journal = {Physical Review B},
  volume  = {75},
  number  = {7},
  pages   = {075119},
  year    = {2007},
  doi     = {10.1103/PhysRevB.75.075119},
  url     = {https://doi.org/10.1103/PhysRevB.75.075119}
}

@article{Liberal2017c,
  author  = {Liberal, Iñigo and Li, Yue and Engheta, Nader},
  title   = {Magnetic field concentration assisted by epsilon-near-zero media},
  journal = {Philosophical Transactions of the Royal Society A: Mathematical, Physical and Engineering Sciences},
  volume  = {375},
  number  = {2090},
  pages   = {20160059},
  year    = {2017},
  doi     = {10.1098/rsta.2016.0059},
  url     = {https://doi.org/10.1098/rsta.2016.0059}
}

@article{Panahpour2025,
  author  = {Panahpour, Ali and Kelavuori, Jussi and Huttunen, Mikko},
  title   = {Purcell Effect in Epsilon-Near-Zero Microcavities},
  journal = {ACS Omega},
  volume  = {10},
  number  = {38},
  pages   = {44683--44692},
  year    = {2025},
}

@article{Kelavuori2024,
  author  = {Kelavuori, J. and Panahpour, A. and Huttunen, M. J.},
  title   = {Dispersion-induced Q-factor enhancement in waveguide-coupled surface lattice resonances},
  journal = {Physical Review B},
  year    = {2024},
  volume  = {110},
  number  = {19},
  pages   = {195422},
  doi     = {10.1103/PhysRevB.110.195422}
}

@book{Bohren2008,
  author    = {Bohren, Craig F. and Huffman, Donald R.},
  title     = {Absorption and Scattering of Light by Small Particles},
  publisher = {John Wiley \& Sons},
  year      = {2008},
  edition   = {1},
  address   = {New York},
  isbn      = {978-0471293408}
}

@article{Feng2016,
  author  = {Feng, Tianhua and Xu, Yi and Liang, Zixian and Zhang, Wei},
  title   = {All-dielectric hollow nanodisk for tailoring magnetic dipole emission},
  journal = {Optics Letters},
  volume  = {41},
  number  = {21},
  pages   = {5011--5014},
  year    = {2016},
  doi     = {10.1364/OL.41.005011},
  url     = {https://doi.org/10.1364/OL.41.005011}
}

@article{Vaskin2019,
  author  = {Vaskin, Aleksandr and Mashhadi, Soheila and Steinert, Michael and Chong, Katie E. and Keene, David and Nanz, Stefan and Abass, A. and Rusak, E. and Choi, D.Y. and Fernandez-Corbaton, I. and Pertsch, T},
  title   = {Manipulation of magnetic dipole emission from Eu\textsuperscript{3+} with Mie-resonant dielectric metasurfaces},
  journal = {Nano Letters},
  volume  = {19},
  number  = {2},
  pages   = {1015--1022},
  year    = {2019},
  doi     = {10.1021/acs.nanolett.8b04400},
  url     = {https://doi.org/10.1021/acs.nanolett.8b04400}
}

@article{Feng2017,
  author  = {Feng, Tianhua and Xu, Yi and Zhang, Wei and Miroshnichenko, Andrey E.},
  title   = {Ideal magnetic dipole scattering},
  journal = {Physical Review Letters},
  volume  = {118},
  number  = {17},
  pages   = {173901},
  year    = {2017},
  doi     = {10.1103/PhysRevLett.118.173901},
  url     = {https://doi.org/10.1103/PhysRevLett.118.173901}
}

@article{Zhang2019,
  author  = {Zhang, Yao and Yue, Peng and Liu, Jun-Yan and Geng, Wei and Bai, Ya-Ting and Liu, Shao-Ding},
  title   = {Ideal magnetic dipole resonances with metal-dielectric-metal hybridized nanodisks},
  journal = {Optics Express},
  volume  = {27},
  number  = {11},
  pages   = {16143--16155},
  year    = {2019},
  doi     = {10.1364/OE.27.016143},
  url     = {https://doi.org/10.1364/OE.27.016143}
}

@article{Baranov2017,
  author  = {Baranov, Denis G. and Savelev, Roman S. and Li, Sergey V. and Krasnok, Alexander E. and Alù, Andrea},
  title   = {Modifying magnetic dipole spontaneous emission with nanophotonic structures},
  journal = {Laser \& Photonics Reviews},
  volume  = {11},
  number  = {3},
  pages   = {1600268},
  year    = {2017},
}

@article{Carletti2015,
  author  = {Carletti, Luca and Locatelli, Andrea and Stepanenko, O. and Leo, G. and De Angelis, Costantino},
  title   = {Enhanced second-harmonic generation from magnetic resonance in AlGaAs nanoantennas},
  journal = {Optics Express},
  volume  = {23},
  number  = {20},
  pages   = {26544--26550},
  year    = {2015},
  doi     = {10.1364/OE.23.026544},
  url     = {https://doi.org/10.1364/OE.23.026544}
}

@article{Obydennov2021,
  author  = {Obydennov, Dmitry V. and Shilkin, Daniil A. and Elyas, Ekaterina I. and Yaroshenko, Vitaly V. and Kudryavtsev, Oleg S. and Zuev, Dmitry A. and Lyubin, Evgeny V. and Ekimov, Evgeny A. and Vlasov, Igor I. and Fedyanin, Andrey A.},
  title   = {Spontaneous light emission assisted by Mie resonances in diamond nanoparticles},
  journal = {Nano Letters},
  volume  = {21},
  number  = {23},
  pages   = {10127--10132},
  year    = {2021},
  doi     = {10.1021/acs.nanolett.1c03650},
  url     = {https://doi.org/10.1021/acs.nanolett.1c03650}
}

@article{Dmitriev2016,
  author  = {Dmitriev, Pavel A. and Baranov, Denis G. and Milichko, Valentin A. and Makarov, Sergey V. and Mukhin, Ivan S. and Samusev, Anton K. and Krasnok, Alexander E. and Belov, Pavel A. and Kivshar, Yuri S.},
  title   = {Resonant Raman scattering from silicon nanoparticles enhanced by magnetic response},
  journal = {Nanoscale},
  volume  = {8},
  number  = {18},
  pages   = {9721--9726},
  year    = {2016},
  doi     = {10.1039/C6NR00878F},
  url     = {https://doi.org/10.1039/C6NR00878F}
}

@article{Kruk2017,
  author  = {Kruk, Sergey and Kivshar, Yuri},
  title   = {Functional meta-optics and nanophotonics governed by Mie resonances},
  journal = {ACS Photonics},
  volume  = {4},
  number  = {11},
  pages   = {2638--2649},
  year    = {2017},
  doi     = {10.1021/acsphotonics.7b01038},
  url     = {https://doi.org/10.1021/acsphotonics.7b01038}
}

@article{Tagviashvili2010,
  author  = {Tagviashvili, M.},
  title   = {$\varepsilon \rightarrow 0$ limits in the Mie-scattering theory},
  journal = {Physical Review A},
  volume  = {81},
  number  = {4},
  pages   = {045802},
  year    = {2010},
  doi     = {10.1103/PhysRevA.81.045802},
  url     = {https://doi.org/10.1103/PhysRevA.81.045802}
}

@article{Silveirinha2006,
  title={Design of matched zero-index metamaterials using nonmagnetic inclusions in epsilon-near-zero media},
  author={Silveirinha, M. G. and Engheta, N.},
  journal={Physical Review Letters},
  volume={97},
  number={15},
 pages={157403},
 year={2006},
  publisher={APS}
}

@INBOOK{Novotny2012,
  author         = {Novotny, Lukas and Bert Hecht},
  booktitle      = {Principles of nano-optics},
  publisher      = {Cambridge university press},
  address        = {},
  title          = {},
  edition        = {1},
  year           = {2012},

}

@article{Qiao2011,
  author  = {Qiao, Peng-fei and Sha, Wei EI and Choy, Wallace C. H. and Chew, Weng Cho},
  title   = {Systematic study of spontaneous emission in a two-dimensional arbitrary inhomogeneous environment},
  journal = {Physical Review A},
  volume  = {83},
  number  = {4},
  pages   = {043824},
  year    = {2011},
  doi     = {10.1103/PhysRevA.83.043824},
  url     = {https://doi.org/10.1103/PhysRevA.83.043824}
}

@ARTICLE{Lobet2020,
   author       = "Lobet, M. and Liberal, I. and Knall, E.N. and Alam, M.Z. and Reshef, O. and Boyd, R.W. and Engheta, N. and Mazur, E.",
   title        = {Fundamental radiative processes in near-zero-index media of various dimensionalities},
   year         = "2020",
   journal      = "ACS Photonics",
   volume       = "7",
   number       = "8",
   pages        = "1965-1970",
   doi          = {10.1021/acsphotonics.0c00782}
}

@article{Malitson1965,
  author  = {Malitson, I. H.},
  title   = {Interspecimen comparison of the refractive index of fused silica},
  journal = {Journal of the Optical Society of America},
  year    = {1965},
  volume  = {55},
  number  = {10},
  pages   = {1205--1209},
  doi     = {10.1364/JOSA.55.001205}
}

@article{Beliaev2022,
  author  = {Beliaev, L. Y. and Shkondin, E. and Lavrinenko, A. V. and Takayama, O.},
  title   = {Optical, structural and composition properties of silicon nitride films deposited by reactive radio-frequency sputtering, low pressure and plasma-enhanced chemical vapor deposition},
  journal = {Thin Solid Films},
  year    = {2022},
  volume  = {763},
  pages   = {139568},
  doi     = {10.1016/j.tsf.2022.139568}
}

@article{Polyanskiy2024,
  author  = {Polyanskiy, Mikhail N.},
  title   = {Refractiveindex.info database of optical constants},
  journal = {Scientific Data},
  year    = {2024},
  volume  = {11},
  number  = {1},
  pages   = {94},
  doi     = {10.1038/s41597-024-02981-y}
}

@article{Chremmos2015,
  author    = {Ioannis D. Chremmos and Efthymios Kallos and Melpomeni Giamalaki and Vassilios Yannopapas and Emmanuel Paspalakis},
  title     = {Effective medium theory for two-dimensional non-magnetic metamaterial lattices up to quadrupole expansions},
  journal   = {Journal of Optics},
  year      = {2015},
  volume    = {17},
  number    = {7},
  pages     = {075102},
  doi       = {10.1088/2040-8978/17/7/075102}
}

@article{Wang2025,
  author    = {Wang, Z. and Lin, R. and Yao, J. and Tsai, D. P.},
  title     = {All-dielectric nonlinear metasurface: from visible to vacuum ultraviolet},
  journal   = {npj Nanophotonics},
  year      = {2025},
  volume    = {2},
  number    = {1},
  pages     = {4},
}

@article{Lin2025Nonlinear,
  author    = {Lin, R. and Yao, J. and Wang, Z. and Chan, C. T. and Tsai, D. P.},
  title     = {Nonlinear Meta-Devices: From Plasmonic to Dielectric},
  journal   = {Engineering},
  year      = {2025},
  volume    = {45},
  pages     = {15--24},
}

@article{Rybin2024,
  author    = {Rybin, M. V. and Kivshar, Yuri},
  title     = {Metaphotonics with subwavelength dielectric resonators},
  journal   = {npj Nanophotonics},
  year      = {2024},
  volume    = {1},
  number    = {1},
  pages     = {43},
}

@article{Kivshar2018AllDielectric,
  author    = {Kivshar, Yuri},
  title     = {All-dielectric meta-optics and non-linear nanophotonics},
  journal   = {National Science Review},
  year      = {2018},
  volume    = {5},
  number    = {2},
  pages     = {144--158}
}

@article{Sugimoto2021,
  author    = {Sugimoto, H. and Fujii, M.},
  title     = {Magnetic Purcell enhancement by magnetic quadrupole resonance of dielectric nanosphere antenna},
  journal   = {ACS Photonics},
  year      = {2021},
  volume    = {8},
  number    = {6},
  pages     = {1794--1800}
}

@article{Liu2020MultipoleMultimode,
  author    = {Liu, T. and Xu, R. and Yu, P. and Wang, Z. and Takahara, J.},
  title     = {Multipole and multimode engineering in Mie resonance-based metastructures},
  journal   = {Nanophotonics},
  year      = {2020},
  volume    = {9},
  number    = {5},
  pages     = {1115--1137}
}

@article{Babicheva2021,
  author    = {Babicheva, V. E. and Evlyukhin, A. B.},
  title     = {Multipole lattice effects in high refractive index metasurfaces},
  journal   = {Journal of Applied Physics},
  year      = {2021},
  volume    = {129},
  number    = {4}
}

@article{Liu2020Dielectric,
  author    = {Liu, W. and Li, Z. and Cheng, H. and Chen, S.},
  title     = {Dielectric resonance-based optical metasurfaces: from fundamentals to applications},
  journal   = {iScience},
  year      = {2020},
  volume    = {23},
  number    = {12}
}

@article{Wang2025encryption,
  author    = {Wang, Y. and Gu, M. and Tian, Y. and Li, C. and Jin, Y. and Wang, L. and Jing, X.},
  title     = {Review on All-Dielectric Metasurface Encryption Technology},
  journal   = {Defence Technology},
  year      = {2025}
}

@article{Ha2018,
  author    = {Ha, S. T. and Fu, Y. H. and Emani, N. K. and Pan, Z. and Bakker, R. M. and Liu, Z. and Haffner, C. and Chew, W. C. and Shalaev, V. M. and Boltasseva, A. and Yang, J. K. W. and Paniagua-Domínguez, R. and Kuznetsov, A. I.},
  title     = {Single and multi-mode directional lasing from arrays of dielectric nanoresonators},
  journal   = {Nature Nanotechnology},
  year      = {2018},
  volume    = {13},
  pages     = {1042--1047}
}

@article{Li2019Symmetry,
  author  = {Li, S. and Zhou, C. and Liu, T. and Xiao, S.},
  title   = {Symmetry-protected bound states in the continuum supported by all-dielectric metasurfaces},
  journal = {Physical Review A},
  year    = {2019},
  volume  = {100},
  number  = {6},
  pages   = {063803}
}

@article{Chern2023,
  author  = {Chern, R.-L. and Yang, H.-C. and Chang, J.-C.},
  title   = {Bound states in the continuum in asymmetric dual-patch metasurfaces},
  journal = {Optics Express},
  year    = {2023},
  volume  = {31},
  number  = {10},
  pages   = {16570--16581}
}

@article{Kasperczyk2015,
  author  = {Kasperczyk, M. and Person, S. and Ananias, D. and Carlos, L. D. and Novotny, L.},
  title   = {Excitation of Magnetic Dipole Transitions at Optical Frequencies},
  journal = {Phys. Rev. Lett.},
  volume  = {114},
  number  = {16},
  pages   = {163903},
  year    = {2015}
}

@article{Das2015_BeamEngineering,
  author  = {Das, Tanya and Iyer, Prasad P. and DeCrescent, Ryan A. and Schuller, Jon A.},
  title   = {Beam engineering for selective and enhanced coupling to multipolar resonances},
  journal = {Phys. Rev. B},
  volume  = {92},
  number  = {24},
  pages   = {241110},
  year    = {2015}
}

@article{Noginova2009,
  author  = {Noginova, N. and Barnakov, Yu. and Li, H. and Noginov, M. A.},
  title   = {Effect of metallic surface on electric dipole and magnetic dipole emission transitions in Eu$^{3+}$–doped polymeric film},
  journal = {Opt. Express},
  volume  = {17},
  number  = {13},
  pages   = {10767--10772},
  year    = {2009}
}

@article{KaraveliZia2011,
  author  = {Karaveli, S. and Zia, R.},
  title   = {Spectral tuning by selective enhancement of electric and magnetic dipole emission},
  journal = {Phys. Rev. Lett.},
  volume  = {106},
  number  = {19},
  pages   = {193004},
  year    = {2011}
}

@article{Aigouy2014,
  author  = {Aigouy, L. and Caz{\'e}, A. and Gredin, P. and Mortier, M. and Carminati, R.},
  title   = {Mapping and quantifying electric and magnetic dipole luminescence at the nanoscale},
  journal = {Phys. Rev. Lett.},
  volume  = {113},
  number  = {7},
  pages   = {076101},
  year    = {2014}
}

@article{Feng2011_MagneticPlasmonic,
  author  = {Feng, Tianhua and Zhou, Ying and Liu, Dahe and Li, Jensen},
  title   = {Controlling Magnetic Dipole Transition with Magnetic Plasmonic Structures},
  journal = {Opt. Lett.},
  volume  = {36},
  number  = {12},
  pages   = {2369--2371},
  year    = {2011},
}

@article{HeinGiessen2013,
  author  = {Hein, Stefan M. and Giessen, Harald},
  title   = {Tailoring Magnetic Dipole Emission with Plasmonic Split-Ring Resonators},
  journal = {Phys. Rev. Lett.},
  volume  = {111},
  number  = {2},
  pages   = {026803},
  year    = {2013},
}

@article{Hussain2015_OptLett,
  author  = {Hussain, Rabia and Kruk, Sergey S. and Bonner, Carl E. and Noginov, Mikhail A. and Staude, Isabelle and Kivshar, Yuri S. and Noginova, Natalia and Neshev, Dragomir N.},
  title   = {Enhancing Eu$^{3+}$ Magnetic Dipole Emission by Resonant Plasmonic Nanostructures},
  journal = {Opt. Lett.},
  volume  = {40},
  number  = {8},
  pages   = {1659--1662},
  year    = {2015},
}

@article{Allayarov2024,
  author  = {Allayarov, I. I. and Kisel, D. A. and Bogdanov, A. A.},
  title   = {Anapole Mechanism of Bound States in the Continuum in Symmetric Dielectric Metasurfaces},
  journal = {Phys. Rev. B},
  volume  = {109},
  number  = {24},
  pages   = {L241405},
  year    = {2024},
  doi     = {10.1103/PhysRevB.109.L241405}
}

@article{Staude2019_MetaOptics,
  author  = {Staude, Isabelle and Pertsch, Thomas and Kivshar, Yuri S.},
  title   = {All-Dielectric Resonant Meta-Optics Lightens Up},
  journal = {Nat. Photonics},
  volume  = {13},
  pages   = {523--525},
  year    = {2019},
}

@article{Babicheva2024,
  author  = {Babicheva, Viktoriia E. and Evlyukhin, Andrey B.},
  title   = {Mie-resonant metaphotonics},
  journal = {Adv. Opt. Photonics},
  volume  = {16},
  number  = {3},
  pages   = {539--620},
  year    = {2024},
}

@article{Hsu2016_BICReview,
  author  = {Hsu, Chia Wei and Zhen, Bo and Stone, A. Douglas and Joannopoulos, John D. and Solja{\v{c}}i{\'c}, Marin},
  title   = {Bound states in the continuum},
  journal = {Nat. Rev. Mater.},
  volume  = {1},
  number  = {9},
  pages   = {16048},
  year    = {2016},
}

@article{Salary2018,
  author  = {Salary, Mohammad Mahdi and Mosallaei, Hossein},
  title   = {Tunable magnetization of infrared epsilon-near-zero media via field-effect modulation},
  journal = {Appl. Phys. Lett.},
  volume  = {112},
  number  = {18},
  pages   = {181104},
  year    = {2018},
}

@article{Liu2019_Photonic,
  author  = {Liu, Na and Zhao, Jia and Du, Liuge and Niu, Chuanning and Lin, Xiao and Wang, Zuojia and Li, Xun},
  title   = {Enhancing the magneto-optical effects in low-biased gyromagnetic media via photonic doping},
  journal = {Opt. Lett.},
  volume  = {44},
  number  = {12},
  pages   = {3000--3003},
  year    = {2019},
}

@article{Nahvi2019,
  author  = {Nahvi, Ehsan and Liberal, I{\~n}igo and Engheta, Nader},
  title   = {Nonperturbative Effective Magnetic Nonlinearity in ENZ Media Doped with Kerr Dielectric Inclusions},
  journal = {ACS Photonics},
  volume  = {6},
  number  = {11},
  pages   = {2823--2831},
  year    = {2019},
}

@article{Shcherbakov2014,
  author  = {Shcherbakov, Maxim R. and Neshev, Dragomir N. and Hopkins, Ben and Shorokhov, Alexander S. and Staude, Isabelle and Melik-Gaykazyan, Elizaveta V. and Decker, Manuel and Ezhov, Alexander A. and Miroshnichenko, Andrey E. and Brener, Igal and Fedyanin, Andrey A. and Kivshar, Yuri S.},
  title   = {Enhanced Third-Harmonic Generation in Silicon Nanoparticles Driven by Magnetic Response},
  journal = {Nano Lett.},
  volume  = {14},
  number  = {11},
  pages   = {6488--6492},
  year    = {2014},
}

@article{Wang2023,
  author  = {Wang, Y. and Xu, P.},
  title   = {Spatial heterogeneity of the doping mode: A potential optical reconfiguration freedom of photonic doping epsilon-near-zero media},
  journal = {Optical Materials},
  volume  = {135},
  pages   = {113300},
  year    = {2023},
}

@article{Luo2018CPA,
  author  = {Luo, J. and Liu, B. and Hang, Z. H. and Lai, Y.},
  title   = {Coherent Perfect Absorption via Photonic Doping of Zero-Index Media},
  journal = {Laser \& Photonics Reviews},
  volume  = {12},
  number  = {8},
  pages   = {1800001},
  year    = {2018},
}

@article{Yan2023,
  author  = {Yan, W. and Zhou, Z. and Li, H. and Li, Y.},
  title   = {Transmission-type photonic doping for high-efficiency epsilon-near-zero supercoupling},
  journal = {Nature Communications},
  volume  = {14},
  number  = {1},
  pages   = {6154},
  year    = {2023},
}

@article{Viitanen2002,
  author  = {Viitanen, A. J. and H{\"a}nninen, I. and Tretyakov, S. A.},
  title   = {Analytical Model for Regular Dense Arrays of Planar Dipole Scatterers},
  journal = {Progress In Electromagnetics Research},
  volume  = {38},
  pages   = {97--110},
  year    = {2002}
}

\medskip

\end{document}


\clearpage

\section*{Section A: Mie Resonances of Cylindrical Inclusions in ENZ Media}

We consider the Mie coefficients corresponding to TE modes of an infinite lossless cylinder:
\begin{equation}
a_n = \frac{x_i J_n(x_i) J_n'(x_e) - x_e J_n(x_e) J_n'(x_i)}{x_i J_n(x_i) {H_n^{(1)}}'(x_e) - x_e H_n^{(1)}(x_e) J_n'(x_i)}.
\end{equation}
The extinction coefficient is proportional to the real part of $a_n$, which can be written in terms of the real numerator ($N$) and complex denominator ($D$) of the coefficient as
\begin{equation}
\Re(a_n) \approx \Re\left( \frac{1}{D} \right) N = \frac{\Re(D) N}{\Re(D)^2 + \Im(D)^2},
\end{equation}
where \(\Re(D)\) and \(\Im(D)\) are the real and imaginary parts of \(D\).
We show that in the ENZ regime, the real part of the Mie coefficients $a_n$ (or $b_n$) behaves like a Lorentzian function normalized to unity at resonance,
\begin{equation}
L(\omega) = \frac{\gamma^2}{\Delta \omega^2 + \gamma^2}.
\end{equation}
In this representation, $\Im(D)$ vanishes at the resonance frequency and thus represents the resonance detuning $\Delta \omega$, while $\Re(D)$ approaches zero as the surrounding refractive index tends to zero, corresponding to the narrowing Lorentzian linewidth $\gamma$.

First, for $n=0$, using the small-argument approximations
\begin{align}
H_0^{(1)}(x_e) &\approx \left( 1 - \frac{x_e^2}{4} \right)
+ i \cdot \frac{2}{\pi} \left( \ln\left( \frac{x_e}{2} \right) + \gamma \right), \\[1em]
H_0^{(1)\prime}(x_e) &\approx -\frac{x_e}{2}
+ i \cdot \left( -\frac{2}{\pi x_e} \right),
\end{align}
the real and imaginary components of the denominator are obtained as
\begin{align}
\Re(D) &= -\frac{x_i x_e}{2} J_0(x_i) - x_e J_0'(x_i) \left( 1 - \frac{x_e^2}{4} \right), \\[1em]
\Im(D) &= -\frac{2 x_i J_0(x_i)}{\pi x_e}
- \frac{2 x_e J_0'(x_i)}{\pi} \left( \ln\left( \frac{x_e}{2} \right) + \gamma \right).
\end{align}
Using the small-argument approximations for the Bessel function,
\begin{align}
J_0(x_e) &\approx 1 - \frac{x_e^2}{4}, \\
J_0'(x_e) &\approx -\frac{x_e}{2},
\end{align}
the simplified expression for the numerator becomes
\begin{equation}
N \approx -\frac{1}{2} x_i x_e J_0(x_i) - x_e J_0'(x_i) \left( 1 - \frac{x_e^2}{4} \right).
\end{equation}
Therefore, when \( x_e \to 0 \), the resonance condition is obtained from the (7) as
$J_0(x_i) = 0$, and for the resonance width we have from (6):
\begin{equation}
\gamma_0 \propto - x_e J_0'(x_i) \left( 1 - \frac{x_e^2}{4} \right),
\end{equation}
which tends to zero as \( x_e \to 0 \). 

For $n>0$, using the small-argument approximations
\begin{equation}
H_n^{(1)}(x_e) \approx \frac{1}{n!}\left( \frac{x_e}{2} \right)^n 
- \frac{1}{(n+1)!} \left( \frac{x_e}{2} \right)^{n+2} 
+ i \cdot \frac{(n-1)!}{\pi} \left( \frac{2}{x_e} \right)^n,
\end{equation}
\begin{equation}
{H_n^{(1)}}'(x_e) \approx \frac{n}{2n!} \left( \frac{x_e}{2} \right)^{n-1} 
- \frac{(n+2)}{2(n+1)!} \left( \frac{x_e}{2} \right)^{n+1} 
- i \cdot \frac{n(n-1)!}{2\pi} \left( \frac{2}{x_e} \right)^{n+1},
\end{equation}
the real and imaginary parts of the denominator are given by
\begin{equation}
\begin{aligned}
\Re(D) =\;
& 
\frac{1}{2n!}\bigl[n x_i J_n(x_i) -  x_e^2 J_n'(x_i)\bigr]\left( \frac{x_e}{2} \right)^{n-1}\\
&\quad 
+\frac{1}{2(n+1)!}\bigl[- (n+2) x_i J_n(x_i) 
+  x_e^2 J_n'(x_i) 
\bigr]\left( \frac{x_e}{2} \right)^{n+1},
\end{aligned}
\end{equation}
\begin{equation}
\Im(D) = 
\frac{(n-1)!}{2\pi} \bigl[ n x_i J_n(x_i) 
- x_e^2 J_n'(x_i) \bigr]\left( \frac{2}{x_e} \right)^{n+1}.
\end{equation}

Now we consider the numerator of the Mie coefficient \( a_n \),
\begin{equation}
N(x_e) = x_i J_n(x_i) J_n'(x_e) - x_e J_n(x_e) J_n'(x_i),
\end{equation}
and expand the Bessel functions up to two terms,
\begin{equation}
J_n(x_e) \approx \frac{1}{n!} \left( \frac{x_e}{2} \right)^n 
- \frac{1}{(n+1)!} \left( \frac{x_e}{2} \right)^{n+2},
\end{equation}
\begin{equation}
J_n'(x_e) \approx \frac{1}{2(n-1)!} \left( \frac{x_e}{2} \right)^{n-1}
- \frac{1}{2(n+1)!} \left( \frac{x_e}{2} \right)^{n+1},
\end{equation}
to get
\begin{equation}
\begin{aligned}
N &\approx \bigl[n x_i J_n(x_i) -  x_e^2 J_n'(x_i)\bigr]\left( \frac{x_e}{2} \right)^{n-1}\\
&\quad -\frac{1}{2(n+1)!}\bigl[x_i J_n(x_i)-x_e^2 J_n'(x_i) \bigr]\left( \frac{x_e}{2} \right)^{n+1}.
\end{aligned}
\end{equation}
We see from (15) that the resonance condition is met by
\begin{equation}
n x_i J_n(x_i) 
- x_e^2 J_n'(x_i) =0,
\end{equation}
which for $x_e \to 0$ reduces to
\begin{equation}
J_n(x_i)=0.
\end{equation}
For the resonance width, we have
\begin{equation}
\gamma_n \propto \frac{-1}{2(n+1)!}\bigl[ (n+2) x_i J_n(x_i) 
-  x_e^2 J_n'(x_i) 
\bigr]\left( \frac{x_e}{2} \right)^{n+1},
\end{equation}
or
\begin{equation}
\gamma_n \propto \frac{1}{2(n+1)!}\bigl[ x_e^2 J_n'(x_i) 
\bigr]\left( \frac{x_e}{2} \right)^{n+1},
\end{equation}
which tends to zero as \( x_e \to 0 \). 

In the case where the electric field is polarized parallel to the cylinder axis (TM polarization), the extinction efficiency is again proportional to the real part of the scattering coefficients $b_n$, expressed as
\begin{equation}
b_n = \frac{x_e J_n(x_i) J_n'(x_e) - x_i J_n(x_e) J_n'(x_i)}{x_e J_n(x_i) {H_n^{(1)}}'(x_e) - x_i H_n^{(1)}(x_e) J_n'(x_i)}.
\end{equation}
Using the small-argument expansions for $n=0$,
\begin{align}
H_0^{(1)}(x_e) &\approx \left( 1 - \frac{x_e^2}{4} \right)
+ i \cdot \frac{2}{\pi} \left( \ln\left( \frac{x_e}{2} \right) + \gamma \right), \\[1em]
H_0^{(1)\prime}(x_e) &\approx -\frac{x_e}{2}
+ i \cdot \left( -\frac{2}{\pi x_e} \right),
\end{align}
\begin{align}
J_0(x_e) &\approx 1 - \frac{x_e^2}{4}, \\
J_0'(x_e) &\approx -\frac{x_e}{2},
\end{align}
the real and imaginary parts of the denominator become
\begin{align}
\Re(D) &= -\frac{x_e^2}{2} J_0(x_i) - x_i J_0'(x_i) \left( 1 - \frac{x_e^2}{4} \right), \\[1em]
\Im(D) &= -\frac{2}{\pi} J_0(x_i)
- \frac{2 x_i}{\pi} J_0'(x_i) \left( \ln\left( \frac{x_e}{2} \right) + \gamma \right),
\end{align}
and the numerator becomes
\begin{equation}
N \approx -\frac{1}{2} x_e^2 J_0(x_i) - x_i J_0'(x_i) \left( 1 - \frac{x_e^2}{4} \right).
\end{equation}
From these expressions we see that to remove the divergence of the denominator, we should have
\begin{equation}
J_0'(x_i)=0,
\end{equation}
and consequently, $\Re(D)$ vanishes when $x_e \to 0$. However, $\Im(D)$ remains non-vanishing in the small-size-parameter limit, preventing the emergence of a resonance condition. Thus, as can also be inferred from Fig.~2b of the main text, the TM\(_0\) mode of the infinite cylinder does not support a true resonance. 

For $n>0$ we use the small-argument approximations
\begin{equation}
H_n^{(1)}(x_e) \approx \frac{1}{n!}\left( \frac{x_e}{2} \right)^n  
+ i \cdot \frac{(n-1)!}{\pi} \left( \frac{2}{x_e} \right)^n,
\end{equation}
\begin{equation}
{H_n^{(1)}}'(x_e) \approx \frac{n}{2n!} \left( \frac{x_e}{2} \right)^{n-1} 
- i \cdot \frac{n(n-1)!}{2\pi} \left( \frac{2}{x_e} \right)^{n+1},
\end{equation}
to obtain the real and imaginary parts of the denominator as 
\begin{equation}
\Re(D_n) = \left( \frac{x_e}{2} \right)^n \cdot \frac{1}{n!}
\left[ \frac{n}{2} J_n(x_i) - x_i J_n'(x_i) \right]
+ \left( \frac{x_e}{2} \right)^{n+2}
\left[ \frac{x_i J_n'(x_i)}{(n+1)!} - \frac{(n + 2)}{2(n + 1)!} J_n(x_i) \right],
\end{equation}
\begin{equation}
\Im(D_n) = - \left( \frac{2}{x_e} \right)^n \cdot \frac{(n - 1)!}{\pi}
\left[ \frac{n}{2} J_n(x_i) + x_i J_n'(x_i)  \right].
\end{equation}
Using only the leading-order terms of the Bessel functions,
\begin{equation}
J_n(x_e) \approx \frac{1}{n!} \left( \frac{x_e}{2} \right)^n,
\quad
J_n'(x_e) \approx \frac{1}{2(n-1)!} \left( \frac{x_e}{2} \right)^{n-1},
\end{equation}
the numerator becomes
\begin{equation}
N \approx
\left( \frac{x_e}{2} \right)^n \cdot \frac{1}{n!} \left[ \frac{n}{2} J_n(x_i) - x_i J_n'(x_i) \right].
\end{equation}
Hence the resonance condition is satisfied when
\begin{equation}
\frac{n}{2} J_n(x_i) + x_i J_n'(x_i) = 0,
\end{equation}
and for the resonance width we have
\begin{equation}
\gamma_n \propto \Re(D_n),
\end{equation}
indicating that the resonance width tends to zero as \( x_e \to 0 \).

\section*{Section B: Mie Resonances of Spherical Nanoparticles in ENZ Media}

To analyze the ENZ limit \( x_e \to 0 \), we use the small-argument expansions of the functions
\begin{equation}
\psi_n(x_e) = \frac{x_e^{n+1}}{(2n+1)!!} + \mathcal{O}(x_e^{n+3}), \quad
\psi_n'(x_e) = \frac{(n+1)x_e^n}{(2n+1)!!} + \mathcal{O}(x_e^{n+2}),
\end{equation}
\begin{equation}
\xi_n(x_e) = \frac{x_e^{n+1}}{(2n+1)!!} + i\left( -\frac{(2n-1)!!}{x_e^n} \right) + \mathcal{O}(x_e^{n+3}) + i\mathcal{O}(x_e^{2-n}),
\end{equation}
\begin{equation}
\xi_n'(x_e) = \frac{(n+1)x_e^n}{(2n+1)!!} + i\left( n \frac{(2n-1)!!}{x_e^{n+1}} \right) + \mathcal{O}(x_e^{n+2}) + i\mathcal{O}(x_e^{1-n}),
\end{equation}
and write the real part of $b_n$ in terms of its numerator ($N$) and denominator ($D$) as
\begin{equation}
\Re(a_n) \approx \Re\left( \frac{1}{D} \right) N = \frac{\Re(D) N}{\Re(D)^2 + \Im(D)^2}.
\end{equation}
Then the numerator $N$, $\Re(D)$ and $\Im(D)$ are calculated as
\begin{equation}
N = \frac{x_e^{n+1}}{(2n+1)!!} \left( (n+1)\psi_n(x_i) - x_i\psi_n'(x_i) \right) + \mathcal{O}(x_e^{n+3}),
\end{equation}
\begin{equation}
\Re(D) = \frac{x_e^{n+1}}{(2n+1)!!} \left( (n+1)\psi_n(x_i) - x_i \psi_n'(x_i) \right)+ \mathcal{O}(x_e^{n+3}),
\end{equation}
\begin{equation}
\Im(D) = \frac{(2n-1)!!}{x_e^n} \left( n\psi_n(x_i) + x_i\psi_n'(x_i) \right)+\mathcal{O}(x_e^{2-n}).
\end{equation}
From these results we see that the dipolar or multipolar magnetic resonances occur when $\Im(D)=0$ or
\begin{equation}
n\psi_n(x_i) + x_i\psi_n'(x_i)=0,
\end{equation}
which is consistent with the blue curves in Fig.~3b of the main text. For the resonance widths we get
\begin{equation}
\gamma_n \propto \frac{x_e^{n+1}}{(2n+1)!!} \left( (n+1)\psi_n(x_i) - x_i \psi_n'(x_i) \right),
\end{equation}
tending to zero as \( x_e \to 0 \).

Similarly, for the $a_n$ coefficients we obtain
\begin{equation}
N = \frac{x_e^n}{(2n+1)!!} \left( (n+1)x_i \psi_n(x_i) - x_e^2 \psi_n'(x_i) \right) + \mathcal{O}(x_e^{n+2}),
\end{equation}
\begin{equation}
\Re(D) = \frac{x_e^n}{(2n+1)!!} \left( (n+1)x_i \psi_n(x_i) - x_e^2 \psi_n'(x_i) \right) + \mathcal{O}(x_e^{n+2}),
\end{equation}
\begin{equation}
\Im(D) = (2n-1)!! \left( \psi_n(x_i)n\frac{x_i}{x_e^{n+1}} + \psi_n'(x_i) \right) + \mathcal{O}(x_e^{1-n}),
\end{equation}
indicating that in the ENZ regime ($x_e \to 0$), the resonance width, proportional to $\Re(D)$, approaches zero. Although the divergence of the first term in $\Im(D)$ suggests that resonances are suppressed in this regime, a more careful analysis reveals that for any arbitrarily small $x_e$, there always exist roots of $\Im(D)$ given by
\begin{equation}
x_i = \frac{\alpha}{2} + \frac{1}{2n} \sqrt{n^2\alpha^2 - 4n x_e^{n+1}},
\end{equation}
located near the zeros $\alpha$ of the function $\psi_n(x)$, which is consistent with the red curves in Fig.~3b.

\subsection*{The roots of $\Im(D)$}

Let \(\alpha\) denote a simple zero of \(\psi_n(x_i)\), that is,
\begin{equation}
\psi_n(\alpha) = 0, \quad \text{and} \quad \psi_n'(\alpha) \neq 0.
\end{equation}
Expanding \(\psi_n(x_i)\) near \(\alpha\) gives
\begin{equation}
\psi_n(x_i) \approx \psi_n'(\alpha) (x_i - \alpha),
\end{equation}
and also
\begin{equation}
\psi_n'(x_i) \approx \psi_n'(\alpha).
\end{equation}
Substituting into the function
\begin{equation}
f(x_i) = \psi_n(x_i) n \frac{x_i}{x_e^{n+1}} + \psi_n'(x_i),
\end{equation}
we find
\begin{equation}
f(x_i) \approx \psi_n'(\alpha) \left( n \frac{x_i (x_i - \alpha)}{x_e^{n+1}} + 1 \right).
\end{equation}
Setting \( f(x_i) = 0 \), we obtain
\begin{equation}
n x_i (x_i - \alpha) + x_e^{n+1} = 0,
\end{equation}
or equivalently,
\begin{equation}
n x_i^2 - n \alpha x_i + x_e^{n+1} = 0,
\end{equation}
which has the solutions
\begin{equation}
x_i = \frac{\alpha}{2} \pm \frac{1}{2n} \sqrt{n^2\alpha^2 - 4n x_e^{n+1}}.
\end{equation}
As \(x_e \to 0\), the square root behaves like
\begin{equation}
\sqrt{n^2\alpha^2 - 4n x_e^{n+1}} \approx n\alpha,
\end{equation}
thus the \(+\) root approaches \(x_i = \alpha\) while the \(-\) root approaches \(x_i = 0\).
Since we seek a root near \(\alpha\), we select the positive-sign root,
\begin{equation}
x_i = \frac{\alpha}{2} + \frac{1}{2n} \sqrt{n^2\alpha^2 - 4n x_e^{n+1}}.
\end{equation}

\section*{Section C: Electric and Magnetic Dipoles, Green Tensor, LDOS, and Purcell Factor}

\subsection*{C.1. Conventions}
We use time-harmonic fields with the phasor convention $e^{-i\omega t}$. 
Free-space constants are $\varepsilon_0$ and $\mu_0$, with $c=1/\sqrt{\varepsilon_0\mu_0}$ and $k_0=\omega/c$.
Throughout, material permeabilities are taken as $\mu(\mathbf r,\omega)=\mu_0$.

\subsection*{C.2. Electric dipole: field and Purcell factor}

Applying the inhomogeneous vector Helmholtz equation
\begin{equation}
\nabla\times\nabla\times \mathbf E(\mathbf r) 
- k_0^2\,\varepsilon(\mathbf r,\omega)\,\mathbf E(\mathbf r)
= i\omega \mu_0\,\mathbf J(\mathbf r),
\end{equation}
the electric dyadic Green tensor $\mathbf G(\mathbf r,\mathbf r';\omega)$ is defined as the solution of
\begin{equation}
\nabla\times\nabla\times \mathbf G(\mathbf r,\mathbf r';\omega)
- k_0^2\,\varepsilon(\mathbf r,\omega)\,\mathbf G(\mathbf r,\mathbf r';\omega)
= \mathbf I\,\delta(\mathbf r-\mathbf r'),
\end{equation}
satisfying the outgoing-wave radiation condition. 
The total field produced by an arbitrary current distribution $\mathbf J(\mathbf r')$ is obtained as
\begin{equation}
\mathbf E(\mathbf r) = i\omega \mu_0 
\int \mathbf G(\mathbf r,\mathbf r';\omega)\,\mathbf J(\mathbf r')\,\mathrm d^3 r'.
\end{equation}

A point electric dipole $\mathbf p$ at $\mathbf r_0$ is represented by a polarization 
$\mathbf P(\mathbf r)=\mathbf p\,\delta(\mathbf r-\mathbf r_0)$, with bound current
\begin{equation}
\mathbf J(\mathbf r) = -\,i\omega\,\mathbf P(\mathbf r)
= -\,i\omega\,\mathbf p\,\delta(\mathbf r-\mathbf r_0).
\end{equation}
Substituting this expression into the integral representation gives
\begin{equation}
\mathbf E(\mathbf r)
= \mu_0 \omega^2\,\mathbf G(\mathbf r,\mathbf r_0;\omega)\,\mathbf p,
\qquad
\mathbf E(\mathbf r_0)
= \mu_0 \omega^2\,\mathbf G(\mathbf r_0,\mathbf r_0;\omega)\,\mathbf p.
\end{equation}
We also write $\mathbf p=p_0\,\mathbf u_p$ with $|\mathbf u_p|=1$.

The normalized electric decay rate (Purcell factor) for a dipole oriented along $\mathbf u_p$ at $\mathbf r_0$ is
\begin{equation}
\frac{\Gamma(\omega)}{\Gamma_0(\omega)}
= \frac{6\pi}{k}\,
\Im\!\left\{
\mathbf u_p\!\cdot\!\mathbf G(\mathbf r_0,\mathbf r_0;\omega)\!\cdot\!\mathbf u_p
\right\},
\qquad k=\frac{\omega}{c}.
\end{equation}
From the above expression for $\mathbf E(\mathbf r_0)$,
\begin{equation}
\mathbf u_p\!\cdot\!\mathbf G(\mathbf r_0,\mathbf r_0;\omega)\!\cdot\!\mathbf u_p
= \frac{\mathbf u_p\!\cdot\!\mathbf E(\mathbf r_0)}{\mu_0 \omega^2\,p_0}.
\end{equation}
Choosing the (arbitrary) source phase $p_0=i|p_0|$ and using
$\Im\{X/i\}=-\Re\{X\}$ for any complex scalar $X$, we obtain
\begin{equation}
\Im\!\left\{
\mathbf u_p\!\cdot\!\mathbf G\!\cdot\!\mathbf u_p
\right\}
= -\,\frac{\Re\!\left\{\mathbf u_p\!\cdot\!\mathbf E(\mathbf r_0)\right\}}
{\mu_0 \omega^2 |p_0|}.
\end{equation}
Substituting into the expression for $\Gamma/\Gamma_0$ yields 
\begin{equation}
{
\frac{\Gamma(\omega)}{\Gamma_0(\omega)}
= -\,\frac{6\pi c}
{\mu_0\,\omega^{3}\,|p_0|}\,
\Re\!\left\{
\mathbf u_p\!\cdot\!\mathbf E(\mathbf r_0,\omega)
\right\}.
}
\end{equation}

\subsection*{C.3. Magnetic dipole: field and Purcell factor}

We include a fictitious magnetic current density $\mathbf M$ to treat magnetic dipoles on the same footing as electric ones (phasor convention $e^{-i\omega t}$):
\begin{align}
\nabla\times \mathbf E &= i\omega \mu_0\,\mathbf H \;-\; \mathbf M, \\
\nabla\times \mathbf H &= \mathbf J \;-\; i\omega \varepsilon_0 \varepsilon(\mathbf r,\omega)\,\mathbf E.
\end{align}

Let $\mathbf G(\mathbf r,\mathbf r';\omega)$ be the electric dyadic Green tensor defined above. One may verify by direct substitution that the magnetic field generated by an arbitrary $\mathbf M$ is
\begin{equation}
\mathbf H(\mathbf r)
= i\omega\,\varepsilon_0 \varepsilon(\mathbf r,\omega)
\int \mathbf G(\mathbf r,\mathbf r';\omega)\,\mathbf M(\mathbf r')\,\mathrm d^3 r'.
\end{equation}
A point-like magnetic dipole $\mathbf m = m_0\,\mathbf u_m$ with $|\mathbf u_m|=1$, located at $\mathbf r_0$ is represented by
\begin{equation}
\mathbf M(\mathbf r) = -\,i\omega\,\mathbf m\,\delta(\mathbf r-\mathbf r_0),
\end{equation}
so that, evaluating at the source point and denoting the local (host) permittivity by
$\varepsilon_b(\omega)=\varepsilon(\mathbf r_0,\omega)$,
\begin{equation}
\mathbf H(\mathbf r_0)
= \varepsilon_0 \varepsilon_b\,\omega^2\,\mathbf G(\mathbf r_0,\mathbf r_0;\omega)\,\mathbf m.
\end{equation}

Let $c_b = 1/\sqrt{\varepsilon_0\varepsilon_b\mu_0}$ be the wave speed and
$k_b=\omega/c_b$ the wavenumber in the host at $\mathbf r_0$. The magnetic Purcell factor (normalized decay rate) is
\begin{equation}
\frac{\Gamma_m(\omega)}{\Gamma_{m,0}(\omega)}
= \frac{6\pi}{k_b}\,
\Im\!\left\{\mathbf u_m\!\cdot\!\mathbf G(\mathbf r_0,\mathbf r_0;\omega)\!\cdot\!\mathbf u_m\right\}.
\end{equation}
Projecting along $\mathbf u_m$ gives
\begin{equation}
\mathbf u_m\!\cdot\!\mathbf G(\mathbf r_0,\mathbf r_0;\omega)\!\cdot\!\mathbf u_m
= \frac{\mathbf u_m\!\cdot\!\mathbf H(\mathbf r_0)}{\varepsilon_0 \varepsilon_b\,\omega^2\,m_0}.
\end{equation}
Choosing the dipole phase $m_0=i|m_0|$ (so that $\Im\{X/i\}=-\Re\{X\}$) yields
\begin{equation}
{\
\frac{\Gamma_m(\omega)}{\Gamma_{m,0}(\omega)}
= -\,\frac{6\pi\,c_b}{\varepsilon_0 \varepsilon_b\,\omega^{3}\,|m_0|}\,
\Re\!\left\{\mathbf u_m\!\cdot\!\mathbf H(\mathbf r_0,\omega)\right\}.\
}
\end{equation}

\section*{Section D: Electric Purcell enhancement in doped ENZ Bragg cavities excited by TM-polarized waves}

The electric Purcell enhancement in NC-doped Bragg cavities can be investigated for different orders of NZI TM-polarized modes by exciting the structure with electric line currents placed at the electric hot spots of the modes. It should be noted that the use of electric or virtual magnetic line currents is purely a computational tool to excite the corresponding TM or TE modes; in practice, the TE- and TM-polarized modes can be excited by plane-wave illumination with the appropriate polarization. Figure~S1 presents the corresponding calculations for the same cavity configuration as in Fig.~14 in the article, but with a reduced width of $W=12.5~\mu\mathrm{m}$ and with electric rather than magnetic line-current excitation. The resulting electric PF spectrum, shown in Fig.~S1a, exhibits a series of sharp resonances corresponding to different orders of NZI TM-like modes. The field distributions for the lowest-order mode at $\lambda=1399.8~\mathrm{nm}$ are shown in Figs.~S1b and S1c, illustrating the magnitudes of the electric field ($|E|$) and magnetic field ($|H|$), respectively. Both profiles demonstrate strong field localization within the NCs, with the inset zoom-ups highlighting the distinct electric and magnetic hot spots of the resonant TM mode.

\setcounter{figure}{0}
\renewcommand{\thefigure}{S\arabic{figure}}

\begin{figure}[ht]
\includegraphics[scale=0.75]{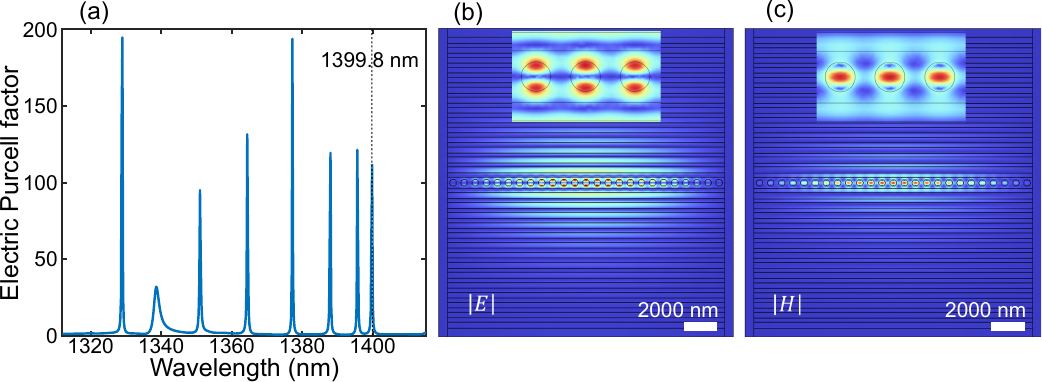}
\caption{\label{fig:S1} 
Electric Purcell enhancement in a Bragg cavity with $N=14$ cladding-layer pairs and width $W=12.5~\mu\mathrm{m}$, loaded with an array of 25 silicon NCs of radius $R=150~\mathrm{nm}$ and period $p=500~\mathrm{nm}$. 
(a) Electric Purcell factor spectrum 
(b) Electric field distribution $|E|$ and (c) magnetic field distribution $|H|$ for the lowest-order TM-like mode at $\lambda=1399.8~\mathrm{nm}$. Insets highlight the strong field localization within the NCs.}
\end{figure}
